\documentclass{article}
\usepackage{bm}
\usepackage{float}
\usepackage{lineno}
\usepackage{natbib}
\usepackage{caption}
\usepackage{amsmath}
\usepackage{hyperref}
\usepackage{setspace}
\usepackage{amsfonts}
\usepackage{graphicx}
\usepackage{csvsimple}
\usepackage[margin=1in]{geometry}
\usepackage{xcolor}
\usepackage{setspace}

\hypersetup{
    colorlinks=true,
    linkcolor=blue,
    citecolor=blue,
    urlcolor=blue,
}

\let\oldequation\equation
\let\oldendequation\endequation
\renewenvironment{equation}
  {\linenomathNonumbers\oldequation}
  {\oldendequation\endlinenomath}

\newcommand{\norm}[1]{\left\lVert{#1}\right\rVert}

\begin{document}

\begin{flushleft}
    \textbf{Reduced-order modeling for complex 3D seismic wave propagation} \\
    John M. Rekoske$^1$, Dave A. May$^1$, Alice-Agnes Gabriel$^{1,2}$ \\
    $^1$Scripps Institution of Oceanography, University of California, San Diego,
    9500 Gilman Drive, La Jolla, CA, USA \\
    $^2$Department of Earth and Environmental Sciences, Ludwig-Maximillians-Universität München, Munich, Germany
\end{flushleft}

\section*{Summary}
Elastodynamic Green’s functions are an essential ingredient in seismology
as they form the connection between direct observations of seismic waves
and the earthquake source. They are also fundamental to various seismological
techniques including physics-based ground motion prediction and kinematic or
dynamic source inversions. In regions with established 3D models of the
Earth's elastic structure, such as southern California, 3D Green's
functions can be computed using numerical simulations of seismic wave
propagation. However, such simulations are computationally expensive
which poses challenges for real-time ground motion prediction
and uncertainty quantification in source inversions.
In this study, we address these challenges by
using a reduced-order model (ROM) approach that enables the rapid evaluation of
approximate Green’s functions.
The ROM technique developed approximates three-component time-dependent surface velocity wavefields obtained from numerical simulations of seismic wave propagation of an arbitrary moment tensor source.
We apply our ROM approach to a 50 km $\times$ 40 km area in the greater Los Angeles area accounting for topography, site effects, 3D subsurface velocity structure, and viscoelastic attenuation.
The ROM constructed for this region enables rapid computation (0.001 CPU hours)
of complete, high-resolution (500 m spacing), 0.5 Hz surface
velocity wavefields that are accurate for a shortest wavelength of 1.0 km
for a single moment tensor source.
Using leave-one-out cross validation,
we measure the accuracy of our Green’s functions for the CVM-S velocity
model in both the time-domain and frequency-domain.
Averaged across all sources, receivers, and time steps, the error in the rapid
seismograms is less than 0.01 cm/s.
We demonstrate that the ROM can accurately and rapidly reproduce simulated seismograms for
generalized moment tensor sources in our region, as well as kinematic
sources by using a finite fault model of the 1987 $M_W$ 5.9 Whittier
Narrows earthquake as an example.
We envision that our rapid,
approximate Green’s functions will be useful for constructing rapid ground motion synthetics with high
spatial resolution, and to improve the uncertainty quantification in
earthquake source inversions.

\section*{Keywords}
Machine learning; computational seismology; earthquake ground motions; wave propagation.

\newpage
\section{Introduction}
Elastodynamic Green's functions are essential tools that
seismologists use to predict ground motions from future earthquakes.
They are also an important ingredient for solving
inverse problems to understand the earthquake source, assuming a
seismic velocity model describing the Earth's structure. Mathematically, the Green's function,
sometimes described in terms of the strain Green's tensor (SGT),
represents the seismic impulse response of a medium \cite[e.g.,][]{zhao2006strain}. It characterizes
the entire wave propagation effects between a source location
and a receiver location \citep{akiQuantitativeSeismology2002}.
One approach to obtain Green's functions is to derive empirical Green's functions (EGFs)
from real seismograms that record seismic events sufficiently similar in location and mechanism to the analyzed earthquake.
However, this approach requires that suitable repeated
events have occurred in the area of
interest \citep[e.g.,][]{hough1998aftershock,hartzell1978earthquake,
abercrombie2015investigating,hough1997empirical,yang2020processing}.
Once obtained, EGFs can be used
to accurately simulate ground motions for future events or to invert for
earthquake source parameters \citep[e.g.,][]{dreger1994empirical,liu2004new}. In contrast to EGFs,
theoretical Green's functions can be calculated by assuming
a (simple) seismic velocity model that describes the Earth's structure.
Theoretical (analytically and numerically derived) Green's functions can be rapidly calculated
for Earth models assuming a 1D
or axisymmetric structure, where the seismic
velocities vary only with depth,
such as through InstaSeis \citep{van2015instaseis,nissen2014axisem}.
However, once lateral variations in the Earth's structure,
or topography, are introduced, there exist no analytical solutions, and numerical simulations to compute Green's functions become computationally more demanding.
Modern crustal seismic velocity models often contain
significant lateral changes in velocity, such as velocity
models describing southern California basins \citep[e.g.,][]{shaw2015unified,magistrale2000scec,
lin2007three,li2023shear}
or subduction zone settings \citep[e.g.,][]{koketsu2008progress,fujiwara2009study,stephenson2017p}.
In such models, 3D Green's functions can deviate from
the ones derived from 1D reference models due to basin
amplification and site effects \citep[e.g.,][]{moschetti2024basin,
moschetti2021seismic,rekoske2022basin,frankel2009sedimentary,olsen2000site}.
Accounting for these effects is crucial for
accurate ground motion prediction and the resulting seismic hazard
estimates.

Numerical simulations of seismic wave propagation remain computationally
expensive and impractical for real-time problems
despite substantial progress in the
underlying numerical methods \citep[e.g.,][]{
komatitsch1998spectral,komatitsch1999introduction,van2020accelerating,
cui2010scalable,rodgers2018broadband,heinecke2014petascale}
and increasing computing power.
Reaching regional-scale, high-frequency simulations which are necessary
for ground motion modeling and probabilistic seismic hazard assessment
can require up to $10^3-10^4$ CPUh on the world's most powerful
supercomputers for just a single earthquake \citep[e.g.,][]{rodgers2020regional,
hu20220,taufiqurrahman2022broadband,graves2011cybershake}.
This computational expense challenges applications in real-time and near-real-time
seismology. Rapid efforts to constrain finite fault models
in real-time \citep[e.g.,][]{bose2012real,bose2018finder,crowell2012real,minson2014real,murray2022impact}
for earthquake early warning and to
produce shake maps \citep{wordenShakeMapManualShakeMap2016}
could be improved by using 3D Green's functions that account
for lateral changes in seismic velocities or attenuation.
Additionally, more detailed rupture models could be used to rapidly constrain
the ground motion levels by using physics-informed estimates for the
inferred slip distribution, compared to empirical ground motion models (GMMs)
which average across many earthquake sources \citep[e.g.,][]{campbell2014nga,
abrahamson2014summary,boore2014nga,chiou2014update}.

Efforts to produce rapid 3D Green's functions
or seismic waveforms using interpolation or machine learning approaches
have shown promise and have been improving in complexity the past few years.
These machine learning approaches typically focus on solving
some version of the seismic wave equation and have used various architectures
including neural operators \citep{yang2021seismic,yang2023rapid,lehmann20243d},
generative adversarial networks \citep{shi2024broadband,wang2021seismogen,florez2022data}
and physics-informed neural networks \citep{moseley2020deep,rasht2022physics,smith2020eikonet}.
However, no current method exists that is capable of performing
regional scale physics-based ground motion prediction that incorporates
detailed structural heterogeneities and topographic effects.
Another important consideration is the time to solution of each method,
and it remains unclear which approaches are best suited for
producing rapid, high-resolution seismic wavefields for
a realistic regional model domain.
Instead of solving the wave equation, other approaches have used
random forests and neural networks to create ground motion maps
or ground motion models using simulated data \citep{monterrubio2024machine,withers2020machine}.

Reduced-order models (ROMs) offer an alternative approach by
simplifying computations and reducing the amount of expensive
numerical simulations described by partial differential equations,
referred to as full-order models (FOMs).
ROMs have been extensively applied in other fields of science
and engineering \citep[e.g.,][]{benner2015survey,hesthaven2022reduced,willcox2002balanced},
as well as in seismology \citep{pereyra2008fast,nagata2023seismic,rekoske2023,hawkins2023model,kusanovic2023soil}.

In this work, we develop an approach for producing rapid,
high-resolution wavefields for a 3D seismic velocity model
using an interpolated proper orthogonal decomposition
reduced-order modeling approach.
This approach substantially extends
the method described in \cite{rekoske2023} by including the handling of moving epicentral source locations,
and generating complete 60-second velocity waveforms for a set
of elementary moment tensors.
It provides a flexible approach to rapidly generate
approximate numerical Green's functions, which in turn can be used to
simulate seismograms for complex rupture models.

We apply this approach to produce Green's functions for
earthquake sources near the Whittier Fault Zone, where
damaging earthquakes, including the 1987 Whittier Narrows blind thrust earthquake, have occurred.
Seismic hazard assessments that use a set of fault locations
may not accurately account for the hazard from these types of blind faults
\citep[e.g.,][]{hauksson1988whittier,shaw1999elusive}.
The seismic wavefields from these earthquakes affect a major metropolitan
area, including part of the Los Angeles Basin,
where local amplification effects are significant \citep{
komatitsch2004simulations,field1997nonlinear,moschetti2024basin}.

The organization of this paper is as follows:
Sec.~\ref{sec:methods} describes the procedures for obtaining
the FOM and ROM seismograms, and the methods to quantitatively evaluate the
accuracy of the ROM using leave-one-out cross validation.
Sec.~\ref{sec:results} provides the seismogram results for three
types of seismic sources: (1) an elementary moment tensor from
the \cite{kikuchi1991inversion} basis, (2) a general moment tensor, and
(3) a kinematic rupture model with a heterogeneous slip distribution.
Lastly, Sec.~\ref{sec:discussion} discusses
potential applications of this work for rapid wavefield
predictions, compares our results to other relevant work, and
outlines potential improvements for future research.

\section{Methods}
\label{sec:methods}
\subsection{Full-order model (FOM): Seismic full waveform modeling using
high-performance computing}
\label{sec:fom}
\begin{figure}
    \centering
    \includegraphics[width=\textwidth]{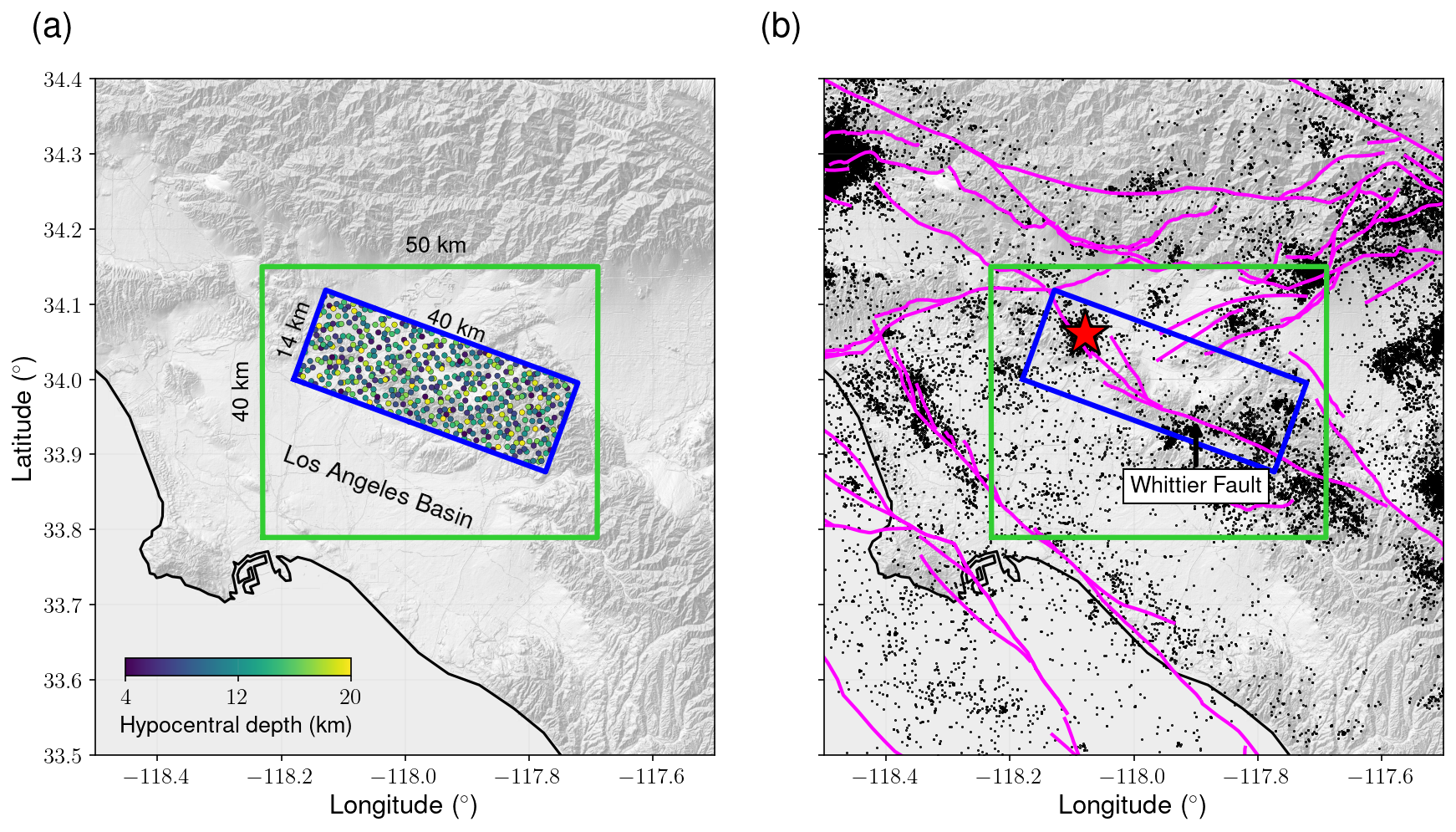}
    \caption{Map of the study area in Southern California used for computing rapid
    seismic wavefields. The source  and receiver areas are indicated by blue and green rectangles, respectively.
    In (a), the earthquake source locations used for the simulations are indicated
    by the colored circles, where the color indicates hypocentral depths ranging from 4 to 20 km. We determine
    the latitude, longitude, and depths of the earthquakes using a pseudorandom Halton sequence.
    In (b), the black dots indicate locations of real earthquakes from the \cite{hauksson2012waveform} catalog, and the magenta lines
    indicate fault traces from the SCEC Community Fault Model \citep{marshall_2023_8327463}. The red star marks
    the epicenter of the 1987 $M_W$ 5.9 Whittier Narrows earthquake, which we use
    as a demonstrator for our finite fault rupture modeling approach.}
    \label{fig:map}
\end{figure}

To generate the synthetic seismogram data for this study,
we perform full waveform modeling for a set of source and receiver
locations in southern California. We here consider
potential earthquakes occurring in a 40 $\times$ 14 $\times$ 16 km
(length $\times$ width $\times$ depth) box which contains a portion of
the Whittier Fault Zone (Fig. \ref{fig:map}).
We will consider hypocentral depths ranging from 4 to 20 km.
In total we consider $n_s = 500$ sources.
The coordinates of the sources are described in Sec.~\ref{sec:rom}.
The receivers are located 1 meter below the Earth's surface on a uniform
grid with 500 m spacing, covering a 50 km by 40 km area,
resulting in a total of $n_r = 8,181$ receivers.
The receivers are located throughout
the densely populated greater Los Angeles area, and cover
a variety of landscapes including the
Los Angeles Basin, the San Gabriel Valley, the Santa Ana Mountains,
and its foothills.

The seismic velocity model describing the
P-wave speed ($V_P$), S-wave speed ($V_S$), and density ($\rho$)
for this study region was defined using the UCVM software \citep{small2017scec}.
We used a particular version of a community velocity model,
CVM-S4.26.M01, that includes a geotechnical layer.
This model was chosen based on results from
\citet{tabordaEvaluationSouthernCalifornia2016},
which indicated that this velocity model performed best for simulating
previous earthquakes in southern California. We use a minimum S-wave speed of 500 m/s and
3D spatially variable quality factors of $Q_S=0.05V_S$ and $Q_P=0.1V_S$
(for $V_S$ measured in m/s) to describe the attenuation of seismic waves
\citep{olsen2003estimation}.

We model visco-elastic seismic wave propagation following the formulation of \cite{kristek2003seismic, moczo2004finite}
in which the constitutive behavior of the domain consists of an elastic body in parallel with $L$ Maxwell (visco-elastic) bodies,
and is governed by the following set of PDEs
\begin{subequations}
\begin{align}
\rho \frac{\partial \bm v}{\partial t} &= \nabla \cdot \bm \sigma, \\
\bm \sigma &= \lambda  \, \text{Tr}(\bm E(\bm u)) \bm I + 2 \mu \bm E(\bm u) - \sum_{l=1}^L \left( \lambda Y_l^\lambda  \, \text{Tr}(\bm \xi^l) \bm I + 2 \mu Y^\mu_l \bm \xi^l \right) + \bm S, \label{eq:pde_sigma}\\
\frac{\partial \bm \xi^l}{\partial t} &= \omega_l \bm E(\bm u) - \omega_l \bm \xi^l ,
\end{align}
\label{eq:pdes}
\end{subequations}
in which $\bm v$ is the velocity, $\bm \sigma$ is the stress, $\bm \xi^l$ are the visco-elastic memory variables for each $l$ mechanism,
$\bm E(\bm u) = \tfrac{1}{2}[ \nabla \bm u + (\nabla \bm u)^T ]$ is the symmetric gradient of the displacement field $\bm u$, $\bm I$ is the identity and
$\text{Tr}(\cdot)$ denotes the trace operator.
The density $\rho$ and Lam{\'e} parameters $\lambda$, $\mu$ are spatially variable and determined using the values of $\rho$, $V_S$, $V_P$ obtained from CVM-S4.26.M01.
The variables $\omega_l$ and $Y_l^\lambda$, $Y_l^\mu$ define the relaxation frequency
and the anelastic analogs of Lam{\'e} parameters for each mechanism $l$.
Following \cite{kaser2007arbitrary} we define the spatially variable anelastic coefficients $Y_l^\lambda$ and $Y_l^\mu$
 using
\begin{equation}
    Y_\ell^\lambda = \left(1 + \frac{2\mu}{\lambda}\right) Y_\ell^P - \frac{2\mu}{\lambda} Y_\ell^S,
    \qquad
    Y_\ell^\mu = Y_\ell^S,
\end{equation}
and
\begin{equation}
    Q_\alpha^{-1} = \sum_{\ell=1}^{L} \frac{\omega_\ell \omega_k + \omega_\ell^2 Q_\alpha^{-1}}{\omega_\ell^2 + \omega_k^2} Y_\ell^\alpha,
    \qquad
    \alpha =P, S,
\end{equation}
together with the assumptions of $Q_S=0.05V_S$ and $Q_P=0.1V_S$
which are defined for the central frequency $\omega_k$. The frequency-independent $Q$ relationships
derived in \cite{olsen2003estimation} describe long-period waves (0 - 0.5 Hz) in the Los
Angeles Basin, thus, we assumed a central frequency $\omega_k$ of 0.2 Hz.

\begin{figure}
    \centering
    \includegraphics[width=\textwidth]{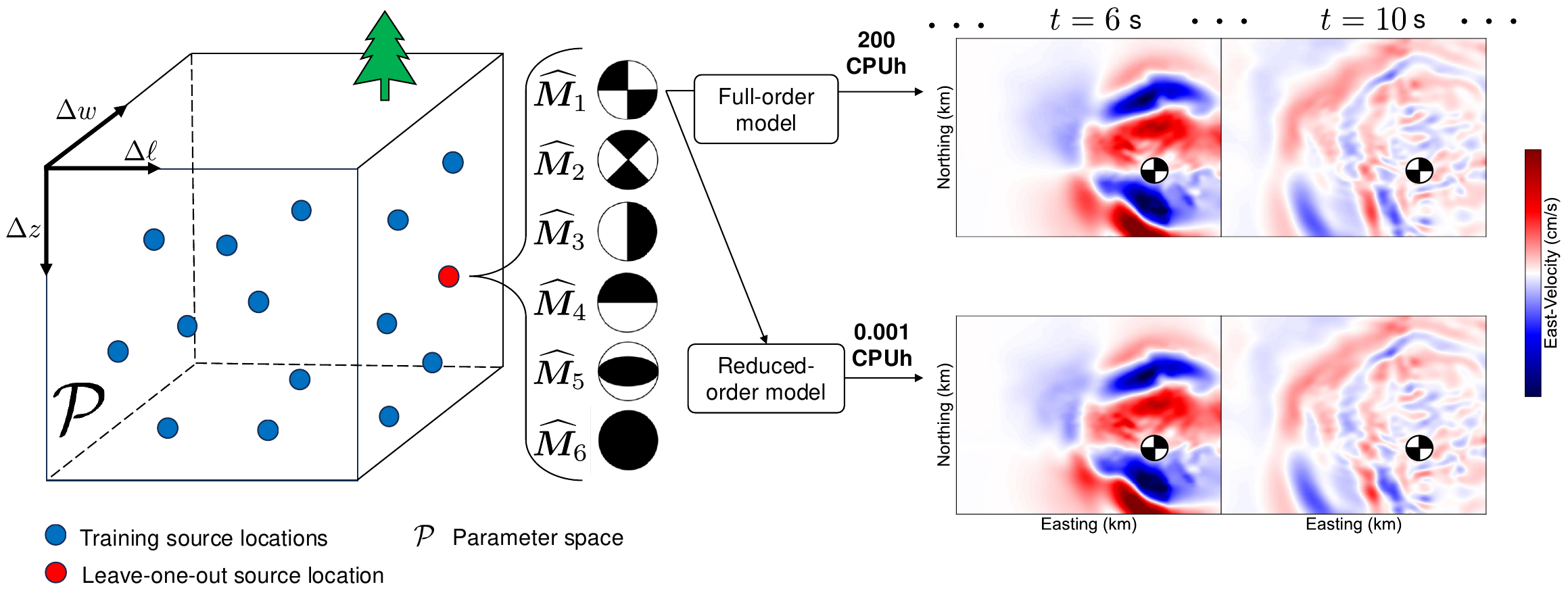}
    \caption{Conceptual illustration of the ROM workflow
    for producing and verifying rapid simulations of seismic wave propagation.
    Within the 3D volume
    defined by $\mathcal{P}$, we simulate forward wave propagation defined by
    the full-order model (FOM) for a set of $n_s$ source locations
    located in the volume, indicated by the colored circles. The source locations are identified by the
    parameters $\Delta\ell$, $\Delta w$, and $\Delta z$ which indicate the distances
    from the sides of the box $\mathcal{P}$ and are determined using a
    pseudorandom Halton sequence. For each source
    location, we run six expensive ($\approx$200 CPUh each) 60-second simulations for each elementary moment tensor
    \citep{kikuchi1991inversion} indicated by the black-and-white beachballs, and store the three-component
    velocity wavefield at the surface. The ROM is a computationally inexpensive
    ($\approx$0.001 CPUh) approximation of the FOM.  We quantify the accuracy of the ROM
    with leave-one-out cross validation.}
    \label{fig:concept}
\end{figure}

In Eq.~\eqref{eq:pde_sigma} the moment tensor seismic source is denoted by $\bm S$.
We assume point sources, thus, it is convenient to decompose $\bm S$ into a time varying component $M(t)$,
a constant tensor $\bar{\bm M}$ and a spatial component resulting in
\begin{equation}
\bm S = M(t) \bar{\bm M} \delta(\bm x - \bm p),
\label{eq:source}
\end{equation}
where $\bm p$ denotes the physical location of the source.
To facilitate representing general moment tensor sources, we exploit the
seismic moment tensor decomposition of \cite{kikuchi1991inversion}
given by
$$
\bar{\bm M} = \sum_{i=1}^6 c_i \widehat{\bm M}_i \,,
$$
in which $\widehat{\bm M}_i$ are referred to as the elementary moment tensors and $c_i$ are
the weights.
The definition of the elementary moment tensors is provided in Appendix \ref{sec:append_mts}
and the beachball plots of the six elementary tensors are shown in Fig.~\ref{fig:concept}.
The moment tensor decomposition can be inverted -- that is, given an arbitrary (symmetric) tensor $\bar{\bm M}$ such as
\begin{equation}
\bar{\bm M}=
    \begin{bmatrix}
        M_{11} & M_{12} & M_{13} \\
        M_{12} & M_{22} & M_{23} \\
        M_{13} & M_{23} & M_{33}
    \end{bmatrix},
\end{equation}
a unique set of weights $c_i$ can be obtained
by solving the following linear system
\begin{equation}
    \begin{bmatrix}
        0 &  1 & 0 & 0 & -1 & 1 \\
        1 &  0 & 0 & 0 &  0 & 0 \\
        0 &  0 & 0 & 1 &  0 & 0 \\
        0 & -1 & 0 & 0 &  0 & 1 \\
        0 &  0 & 1 & 0 &  0 & 0 \\
        0 &  0 & 0 & 0 &  1 & 1
    \end{bmatrix}\begin{bmatrix}
        c_{1} \\ c_{2} \\ c_{3} \\ c_{4} \\ c_{5} \\ c_{6}
    \end{bmatrix}=\begin{bmatrix}
        M_{11} \\ M_{12} \\ M_{13} \\ M_{22} \\ M_{23} \\ M_{33}
    \end{bmatrix}\,.
    \label{eq:elem_weights}
\end{equation}
Any general moment tensor can be reconstructed using a combination
of the six elementary tensors. One advantage of
 this decomposition is that some sources may not require
all six components. For example, only two components are required for pure strike-slip
earthquakes, and only five components are required
for pure double-couple sources \citep{kikuchi1991inversion}.
We note that here, our method slightly differs from CyberShake simulations
\citep[e.g.,][]{graves2011cybershake,milner2021toward,jordan2018cybershake}
where SGTs are computed for only dip-slip and strike-slip components
for assumed fault orientations.
We use the moment rate as a function of time $t$ given by
$\dot{M}(t)=\frac{M_0t}{T^2}$ with $M_0=1.0\times 10^{15}$ N$\,$m and $T=0.34$ s. In the frequency domain,
this source has a Brune-like spectrum corresponding to a stress drop
of approximately 5.0 MPa \citep{brune1970tectonic}.

Lastly we note that Eqs.~\eqref{eq:pdes} are linear in the velocity $\bm v$, stress $\bm \sigma$ and $\bm \xi^l$,
hence can we can apply the principal of linear superposition.
That is, given a set of $k = 1, \dots, N$ source terms $\bm S_k$ of the form shown in Eq.~\eqref{eq:source}, the resulting solutions $\bm v_1, \dots \bm v_N$, $\bm \sigma_1, \dots, \bm \sigma_N$ and $\bm \xi^l_1, \dots, \bm \xi^k_N$
can be linearly combined such that $\bm v = \sum_{k=1}^N \bm v_k$, $\bm \sigma = \sum_{k=1}^N \bm \sigma_k$, $\bm \xi^l = \sum_{k=1}^N \bm \xi^l_k$ is the solution to Eqs.~\eqref{eq:pdes} with $\bm S = \sum_{k=1}^N \bm S_k$.

We solve Eqs.~\eqref{eq:pdes} using SeisSol (\url{www.seissol.org}),
an open-source software for numerical simulation of seismic wave propagation and earthquake
dynamics. SeisSol uses the Arbitrary high-order accurate DERivative Discontinuous Galerkin method (ADER-DG) \citep{KaserDumbser2006,DumbserKaeser2006,kaser2007arbitrary} and end-to-end
optimization for high-performance computing infrastructure \citep{heinecke2014petascale,uphoff2016generating,Uphoff2017,Rettenberger2016}. SeisSol employs fully
unstructured tetrahedral meshes which allow for geometrically complex 3D
geological structures and topography to be accurately resolved.
We use velocity-aware meshing \citep{BreuerHeinecke2022} to create a mesh consisting of
about 3.5 million tetrahedral elements
that accurately resolves frequencies up to 0.5 Hz.
This estimate is based on requiring six elements per shortest wavelength
according to the numerical accuracy analysis
of SeisSol's modal ADER-DG approach performed by \cite{kaserQuantitativeAccuracyAnalysis2008}.
Within the visco-elastic framework used we consider three damping mechanisms ($L=3$)
with relaxation frequencies of $\omega_1 = 0.02$ Hz, $\omega_2 = 0.2$ Hz, and $\omega_3 = 2.0$ Hz.
The value of $\omega_L$ (maximum frequency)
is typically the upper bound for which $Q$ is assumed to be frequency-independent.
We used a value of $\omega_L$ which is slightly higher than the largest frequency resolvable with our mesh to ensure $Q$ is frequency-independent in our band of interest.
The choice of $L=3$ with logarithmically spaced relaxation frequencies was taken as
this configuration is known to be sufficient to achieve less than 5\% error in the frequency band
of interest \citep{emmerich1987incorporation}.

The computational cost per simulation is approximately 192 central processing unit hours (CPUh) per point source location
using a degree four polynomial in space and time.
All numerical simulations were performed on SuperMUC-NG using single
precision.
For each seismic source, we simulate 60 seconds of wave propagation.
For all six elementary tensors and
for all $n_s = 500$ source locations,
the total cost of generating the synthetic data was approximately 576,000
CPUh. The number of receivers is much larger than the number of sources,
thus, we run the simulations in a forward sense and not a reciprocal
sense. From each simulation, we save the 60-second velocity
waveforms, sampled at 10 Hz, for each of the 8,181 receivers.
We postprocess the timeseries data using a linear, fourth-order, Butterworth
low-pass filter at 0.5 Hz. We apply the filter both
forwards and backwards to ensure zero phase shift of the
waveform data.

SeisSol can also compute synthetic seismograms for finite
earthquake sources that are described by a set of point sources, such as
kinematic rupture models.
Examples of these finite sources can be
found at the SRCMOD database \citep{maiSRCMODOnlineDatabase2014}.
When simulating finite sources the moment rate function
at each point source may be variable.
When considering finite sources, we do not apply any upsampling
in space, meaning that we use the same number of moment tensors in our
forward simulations as are provided in the original rupture models. The exact computation
time per forward simulation of a finite source
may slightly vary depending on the source description, though we found that
when using a model composed of 100 point sources, the computation
time was approximately the same at 198 CPUh.

\subsection{Reduced-order model (ROM)}
\label{sec:rom}

The ROM that we present in this study is a computationally inexpensive,
data-driven technique to produce seismic wavefields for earthquakes in our study region.
We build the ROM from the synthetic data (simulated surface velocity waveforms) generated for a set
of $n_s=500$ sources located within the volume $\mathcal{P}$ for the six elementary moment tensors.
We denote each source location by the vector $\bm{p}_i$ given by
\begin{equation}
    \bm{p}_i=\begin{bmatrix}\Delta\ell_i \\ \Delta w_i \\ \Delta z_i\end{bmatrix},
    \qquad
    i=1, \ldots, n_s\,,
\end{equation}
where $\Delta\ell_i, \Delta w_i$, and $\Delta z_i$ denote the three distances
to the sides of the box $\mathcal{P}$ (see Fig. \ref{fig:concept}).
We prescribe the locations of the earthquake sources
in three dimensions using a pseudorandom Halton sequence,
which efficiently fills the entire volume $\mathcal{P}$. The locations of the
sources generated by the Halton sequence
are shown in Fig. \ref{fig:map}.

After performing all the wave propagation simulations
for these sources, as described by the FOM procedure in Sec.~\ref{sec:fom},
we gather all lowpass-filtered receiver timeseries data
into matrices $\bm{D}_{ij}$ where $i=1,\dots, 6$ indicates
the elementary tensor index and $j=1,2,3$ indicates the component
of the wavefield ($j=1$ is the $u$-component,
$j=2$ is the $v$-component, and $j=3$ is the $w$-component).
We denote $\bm{D}_{ij}$ as
\begin{equation}
    \bm{D}_{ij}=\begin{bmatrix}
        \bm{q}_{ij}(\bm{p}_1),  & \bm{q}_{ij}(\bm{p}_2),  & \ldots, & \bm{q}_{ij}(\bm{p}_{n_s})
    \end{bmatrix},
\end{equation}
where $\bm{q}_{ij}$ is a vector containing the surface
velocity timeseries data, produced by the FOM,
for all $n_r=8,181$ receivers and all $n_t=600$ time steps. Thus, each
$\bm{D}_{ij}$ contains $n_r\times n_t=8,181\times600=4,908,600$ rows and
$n_s=500$ columns.
We then compute the singular value decomposition (SVD) of each $\bm{D}_{ij}$,
\begin{equation}
    \bm{D}_{ij}=\bm{U}_{ij}\bm{\Sigma}_{ij}\bm{V}_{ij}^T \,,
    \label{eq:svd}
\end{equation}
where $\bm{U}_{ij}$ contains the left singular vectors (also known
as modes), $\bm{\Sigma}_{ij}$ contains
the singular values, and $\bm{V}_{ij}$ contains the right singular vectors.
We emphasize that our approach considers the SVD for each elementary tensor and velocity component
separately as the total size of the simulated data is approximately
$\approx$177 GB.\footnote{All elements of ${\bm D}_{ij}$ are represented via the \texttt{float} datatype (4 bytes).}
A further computational saving is made by
not computing the SVD directly, but rather computing the eigen decomposition
of $\bm C_{ij} = \bm{D}_{ij}^T\bm{D}_{ij}$. The singular values
$\bm{\Sigma}_{ij}$ are given by the
square root of the eigenvalues of $\bm{C}_{ij}$,
and the right singular vectors, $\bm{V}_{ij}$,
are equal to the eigenvectors of $\bm{C}_{ij}$.
We can obtain the left singular vectors by solving
Eq.~\eqref{eq:svd} for $\bm{U}_{ij}$.

The act of computing the SVD of the data matrices ${\bm D}_{ij}$ defines a set of (discrete) spatial basis functions (the left singular vectors),
along with weights of coefficients (singular values and right singular vectors) which can be used to
reconstruct any column in $\bm D_{ij}$.
This interpretation of the SVD is connected with the term Proper Orthogonal Decomposition (POD) coined
in the seminal model order reduction literature \citep[e.g,][]{berkooz1993proper,bui2003proper,druault2005use}.

After computing the singular values and right singular vectors, we
define the POD coefficients as the matrix $\bm{A}_{ij}$ given by
\begin{equation}
    \bm{A}_{ij}=\bm{V}_{ij}\bm{\Sigma}_{ij}\,.
    \label{eq:pod_coeffs}
\end{equation}
We follow
\cite{rekoske2023} to interpolate the POD coefficients
using radial basis functions (RBFs) \cite[e.g,][]{lazzaro2002radial,audouze2009reduced,xiao2015non}.
That is, the interpolator
function for the $k$-th mode associated with $\bm{A}_{ij}$ is
\begin{equation}
    \tilde{A}_{ijk}(\bm{p})=\sum_{l=1}^{n_s} w_{ijkl}\varphi(|| \bm{p} -\bm{p}_l ||_2)+\sum_{m=1}^{|V|} b_{ijkm}\psi_m(\bm{p}) \,,
    \label{eq:rbf}
\end{equation}
where the function $\varphi(\cdot)$ is an RBF, $w_{ijkl}$ in an entry
of the kernel weight matrix $\bm{W}_{ij}$, $b_{ijkm}$ is an entry of the
polynomial coefficient matrix $\bm{B}_{ij}$, and $\psi_m(\cdot)$ are
monomial terms up to a specified degree $d$ contained in the set $V$
given by
\begin{equation}
    V = \left\{ \Delta\ell^r\Delta w^s \Delta z ^t   \,\big|\,  r+s+t \leq d \,\,\, \text{with} \,\,\, r \ge 0, s \ge 0, t \ge 0  \right\}.
\end{equation}
We determine the coefficients contained in $\bm{W}_{ij}$ and $\bm{B}_{ij}$
by solving the system of linear equations
\begin{subequations}
\begin{align}
    \bm{\Phi W}_{ij}+\bm{\Psi B}_{ij} &= \bm{A}_{ij}\,,  \\
    \bm{\Psi}^T\bm{B}_{ij} &= \bm{0}\,,
\end{align}
    \label{eq:sys}
\end{subequations}
where
\begin{equation}
    \bm{\Phi}=
    \begin{bmatrix}
    \varphi(\norm{\bm{p}_1-\bm{p}_1}_2) & \varphi(\norm{\bm{p}_2-\bm{p}_1}_2) & \ldots & \varphi(\norm{\bm{p}_{n_s}-\bm{p}_1}_2) \\
    \varphi(\norm{\bm{p}_1-\bm{p}_2}_2) & \varphi(\norm{\bm{p}_2-\bm{p}_2}_2) & \ldots & \varphi(\norm{\bm{p}_{n_s}-\bm{p}_2}_2) \\
    \vdots & \vdots & \ddots & \vdots \\
    \varphi(\norm{\bm{p}_1-\bm{p}_{n_s}}_2) & \varphi(\norm{\bm{p}_2-\bm{p}_{n_s}}_2) & \ldots & \varphi(\norm{\bm{p}_{n_s}-\bm{p}_{n_s}}_2) \\
    \end{bmatrix} \,,
\end{equation}
and $\bm{\Psi}$ is the matrix of polynomials up to degree $d$.
For the RBFs $\varphi(\cdot)$, we test four types of kernels:
linear ($\varphi(r)=r$),
thin plate spline ($\varphi(r)=r^2\ln(r)$), cubic
($\varphi(r)=r^3$) and quintic ($\varphi(r)=-r^5$). For each kernel,
we solve the linear system using the minimum polynomial degree,
i.e., $d=0$ for linear, $d=1$ for the thin plate spline and cubic kernels, and $d=2$
for the quintic kernel.

To generate seismic waveforms using the ROM for a new source
location given by $\bm{p}^{\ast}$, for an elementary
moment tensor, we can perform a reconstruction
using the POD basis (i.e. the left singular vectors) via
\begin{equation}
    \tilde{\bm{q}}_{ij}(\bm{p}^{\ast})=\sum_{k=1}^{n_s} \tilde{A}_{ijk}(\bm{p}^{\ast})\bm{u}_{ijk}\,,
    \label{eq:recon_elem}
\end{equation}
where $\bm{u}_{ijk}$ indicates the left singular vector from
the matrix $\bm{U}_{ij}$ for the $k$-th mode. $\tilde{\bm{q}}_{ij}$ then contains
the predicted data for the $i$-th elementary tensor and the
$j$-th wavefield component, and the tilde indicates
that this is an approximation.

To simulate seismic waveforms via the ROM for a source described
by a general moment tensor $\bar{\bm M}$,
we first obtain the
weights $c_i$ by solving Eq.~\eqref{eq:elem_weights}
where $c_i$ is the weight corresponding to the $i$-th elementary tensor,
and then compute
\begin{equation}
    \tilde{\bm{q}}_{j}(\bm{p}^{\ast})=\sum_{i=1}^{6}c_i\tilde{\bm{q}}_{ij}(\bm{p}^{\ast})\,,
    \label{eq:recon_mt}
\end{equation}
where the $\tilde{\bm{q}}_{ij}$ are obtained using Eq. (\ref{eq:recon_elem}).
That is, we weight the ROM solution obtained from each elementary tensor $i$ by $c_i$.
If any of the $c_i$ are equal to zero, then all six components
might not be needed in Eq. (\ref{eq:recon_mt}).

\subsection{Training and validation procedure: leave-one-out cross validation}
\label{sec:validation}
To quantify the accuracy of the rapid seismic wavefields obtained from the ROM,
we use a leave-one-out cross-validation (LOOCV) procedure \citep{rippaAlgorithmSelectingGood1999}.
One advantage of using RBFs as an interpolator function
is that it is relatively cheap to perform this type of analysis
using the \cite{rippaAlgorithmSelectingGood1999} algorithm,
compared to machine learning techniques where re-training
models can be very costly. The \cite{rippaAlgorithmSelectingGood1999}
LOOCV procedure is outlined below:

\begin{enumerate}
\item Write the system in Eq.~\eqref{eq:sys} as
    \begin{equation}
        \begin{pmatrix}
            \bm{\Phi} & \bm{\Psi} \\
            \bm{\Psi}^T & \bm{0}
        \end{pmatrix}
        \begin{pmatrix}
        \bm W_{ij} \\
        \bm B_{ij}
        \end{pmatrix}
=
        \begin{pmatrix}
        \bm A_{ij} \\
        \bm 0
        \end{pmatrix}
        \quad
        \rightarrow
        \quad
         \bm{\Gamma} \bm X_{ij} = \bm F_{ij},
    \end{equation}
    where $\bm X_{ij}$ and $\bm F_{ij}$ represent a matrix of solutions and right-hand sides, respectively.
    We then note that $\bm{\Gamma}$ is identical for every
    elementary moment tensor $i$ and wavefield component $j$.
    Given this, it is efficient to factor $\bm{\Gamma}$ once using a stable decomposition such as the Cholesky decomposition or SVD,
    and then apply the action of the inverse decomposition to obtain each $\bm X_{ij}$.
    \item Re-use the factored form of $\bm{\Gamma}$ to compute $\bm{v}$,
    defined as the first $n_s$ components of the diagonal of $\bm{\Gamma}^{-1}$.
    \item Compute the leave-one-out predictions for the POD coefficients via
    \begin{equation}
    \tilde{A}_{ijk}(\bm{p}_l) = A_{ijkl} - w_{ijkl} \odot \frac{1}{v_{l}},
    \end{equation}
    where $A_{ijkl}$ indicates the POD coefficient from $\bm{A}_{ij}$ that
    corresponds to the $k$-th mode and the $l$-th source location and $\odot$ indicates an element-wise product.
    \item Use Eq. (\ref{eq:recon_elem}) to compute the predicted data
    $\tilde{\bm{q}}_{ij}(\bm{p}_l)$ using the predicted POD coefficients
    $\tilde{A}_{ijk}$ and left singular vectors $\bm{U}_{ij}$.
\end{enumerate}

In addition to the LOOCV error, we also compute the following accuracy metrics for
    the mean absolute velocity error (MAVE), mean peak ground
    velocity error (MPGVE), and the mean spectral error (MSE):
    \begin{align}
        \textrm{MAVE}_{ij}(\bm{p})&=\frac{1}{n_rn_t}\left|\left|\bm{q}_{ij}(\bm{p}) - \tilde{\bm{q}}_{ij}(\bm{p})\right|\right|_1, \\
        \textrm{MPGVE}_{ij}(\bm{p})&=\frac{1}{n_r}\left|\left|\max_t|\bm{q}_{ij}(\bm{p})|-\max_t|\tilde{\bm{q}}_{ij}(\bm{p})|\right|\right|_1, \\
        \textrm{MSE}_{ij}(\bm{p}, f)&=\frac{1}{n_r}\left|\left|\hat{\bm{q}}_{ij}(\bm{p}, f)-\hat{\tilde{\bm{q}}}_{ij}(\bm{p}, f)\right|\right|_1.
    \end{align}
    Here, $\hat{\bm{q}}_{ij}(\bm{p}, f)$ indicates the Fourier transform
    of the FOM velocity timeseries as a function of frequency $f$.
    We compute the FFTs using \texttt{numpy.fft.rfft}, take the absolute
    value, then normalize by multiplying by the sampling interval
    of 0.1 s.

\subsection{Approximate Green's function calculations}
\label{sec:green}
The seismograms that can be computed directly using our ROM correspond
to a particular moment rate function. If we instead had used a
delta function for the source time function, then the seismograms
would be mathematically equivalent to the Green's function;
however, it is often more numerically stable to use a smooth source time
function with a fixed width \citep[e.g.,][]{igel2017computational,van2015instaseis}.
Since it is often required to simulate seismograms for rupture models with different moment rates,
we compute the elastodynamic Green's functions by deconvolving the
prescribed source time function from the seismograms. Following
\cite{akiQuantitativeSeismology2002},
our simulated velocity seismograms can be
expressed as a convolution of the source's moment rate
function and a Green's function,
\begin{equation}
    \bm{q}_{ij}(\bm{p}) = \dot{M} \ast \bm G_{ij}(\bm{p}).
\end{equation}
Here, $\bm G_{ij}(\bm{p})$ contains the time-dependent
Green's functions for all receiver locations for the source
located at $\bm{p}$. Using the convolution theorem, we obtain
\begin{equation}
    \bm G_{ij}(\bm{p})=\mathcal{F}^{-1}\left\{\frac{\mathcal{F}\left\{\bm{q}_{ij}(\bm{p})\right\}}{\mathcal{F}\left\{\dot{M}\right\}}\right\}
\end{equation}
where $\mathcal{F}$ indicates the Fourier transform. Replacing
the FOM seismograms with the ROM-predicted seismograms,
we obtain the approximate Green's functions,
\begin{equation}
    \tilde{\bm G}_{ij}(\bm{p})=\mathcal{F}^{-1}\left\{\frac{\mathcal{F}\left\{\tilde{\bm{q}}_{ij}(\bm{p})\right\}}{\mathcal{F}\left\{\dot{M}\right\}}\right\}.
    \label{eq:approx_green}
\end{equation}
Thus, we can use the computed equivalent Green's functions
to rapidly approximate the simulated data from
finite fault rupture models using
the discrete representation theorem \citep{akiQuantitativeSeismology2002}.
For example, if we have a rupture model consisting of
$n_f$ subfaults, where each subfault may have a different
source time function ($\dot{M}^k$), the general formula for the
obtaining the predicted velocity timeseries for component $j$ is given by
\begin{equation}
    \tilde{\bm{q}}_{j}=\sum_{k=1}^{n_f}\sum_{i=1}^6 c_{i}^k \dot{M}^k \ast \tilde{\bm G}_{ij}(\bm{p}_k),
    \label{eq:recon_ff}
\end{equation}
where $c_{i}^k$ indicates the weight for the $i$-th elementary
moment tensor for the $k$-th subfault.

\section{Results}
\label{sec:results}
\subsection{Singular value spectra of the simulated seismograms}
\begin{figure}
    \centering
    \includegraphics[width=\textwidth]{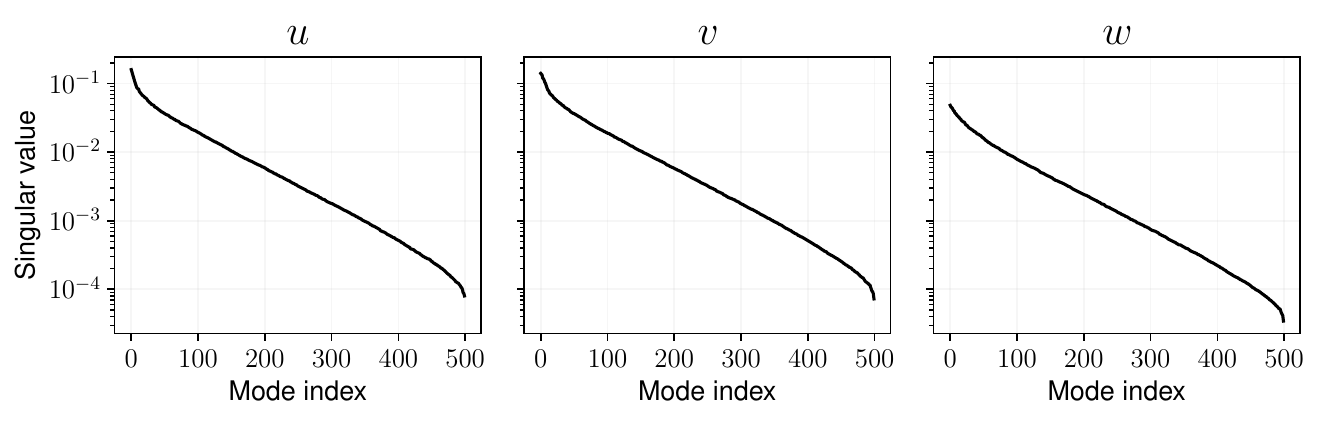}
    \caption{Singular value spectra obtained by taking the SVD of
    the data matrices which contain the FOM seismograms
    corresponding to the east ($u$), north ($v$), and vertical ($w$)
    components of the seismic wavefield, respectively.
    The singular values are shown for the simulated seismograms associated with the
    $\widehat{\bm M}_1$ elementary moment tensor source (see Appendix \ref{sec:append_mts}
    for the definitions of the elementary moment tensors).}
    \label{fig:sigvals}
\end{figure}

Taking the SVD of the data matrices of simulated seismograms allows us to obtain
the singular values which provide important understanding of the properties and structure of
each data matrix. Specifically, the singular values indicate how important the information is
that is contained in each of the modes.
The singular values for the $\widehat{\bm M}_1$
elementary moment tensor (a pure strike-slip fault) associated
with each velocity component are shown in Fig. \ref{fig:sigvals}.
The singular values for each velocity component decrease rapidly
over the first 20 indices, and then decay with an apparent log-linear slope.
We infer that the data matrices have (approximately) full rank, since the singular values over the entire spectrum do not decay exponentially fast.
As expected, the singular values for the $w$-component
(vertical) are generally smaller than the singular
values for the horizontal components due to the smaller amplitudes of the seismic
waves on the vertical component.

\begin{figure}
    \centering
    \includegraphics[width=\textwidth]{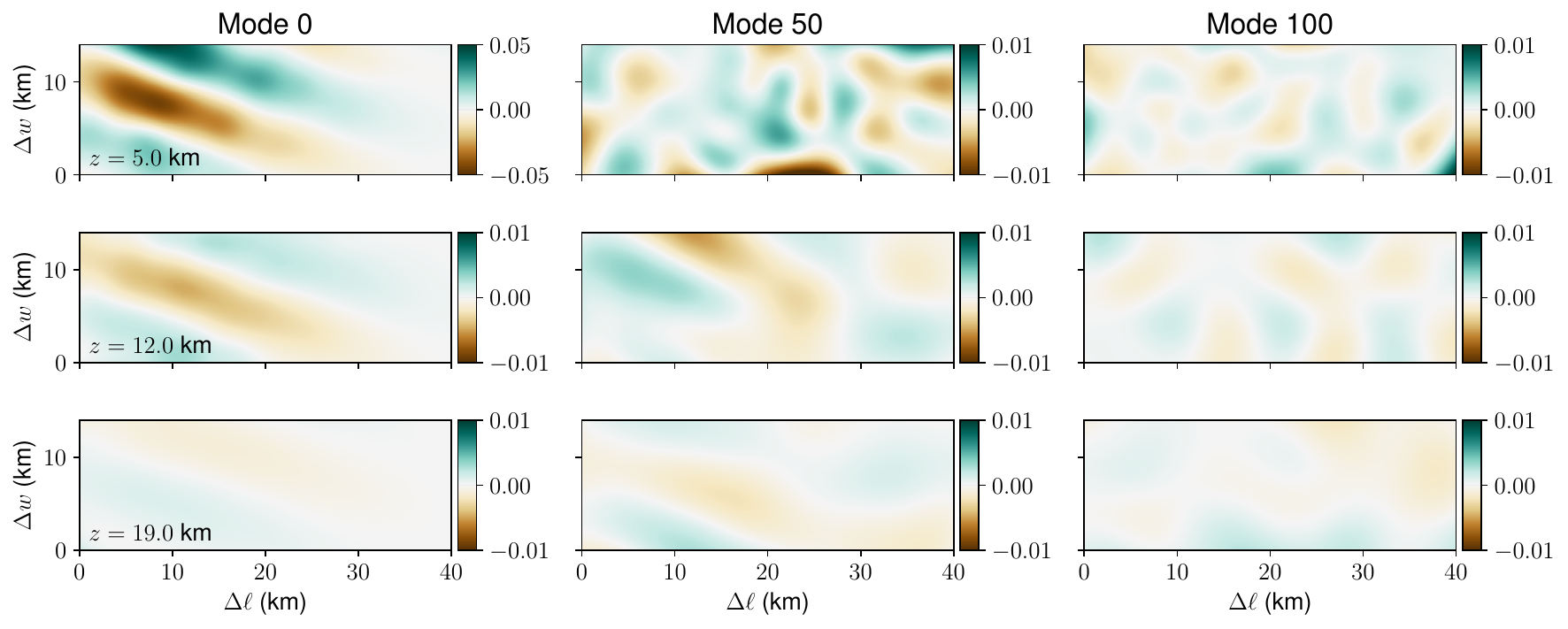}
    \caption{Views of the POD coefficients predicted by the RBFs for three
    different depth slices at $z=(5, 12$, and $19)$ km. The results are shown
    for the quintic RBF kernel for the first elementary moment tensor
    and the east-component of the wavefield. The amplitudes of the POD coefficients are smaller
    for deeper earthquakes and for the higher-order modes
    (note the change in color scale in the top left panel
    which shows the Mode 0 coefficients for the shallow depth).}
    \label{fig:coeffs}
\end{figure}

When plotted in depth slices, the POD coefficients appear as smooth
maps of varying positive and negative amplitudes
that can be accurately interpolated (Fig. \ref{fig:coeffs}).
Following the definition of the POD, we see that the amplitudes of the POD coefficients
decrease as the index of the mode increases due to the ordering
of the singular values. Also, the coefficients
decrease in amplitude for deeper earthquakes, which is also expected
due to the decrease in ground motion amplitudes for deeper earthquakes.

\subsection{FOM and ROM accuracy analysis for an elementary moment
tensor source}
\label{sec:res-m0}
Using the interpolator functions to predict POD coefficients,
we can now inspect how accurately the rapid seismic wavefields produced by the
ROM match our high-resolution FOM simulations.
We first inspect the wavefields that are produced by a point source
$\widehat{\bm M}_1$ elementary moment tensor. We select the results for
the first source location from the Halton sequence as an example.
This source is located at ($\Delta\ell=3.5$ km, $\Delta w=9.9$ km, $\Delta z=12.5$ km).
The ROM wavefield errors are evaluated using the \cite{rippaAlgorithmSelectingGood1999} algorithm
for leave-one-out cross-validation as described in Sec.~\ref{sec:validation}.

\begin{figure}
    \centering
    \includegraphics[width=0.9\textwidth]{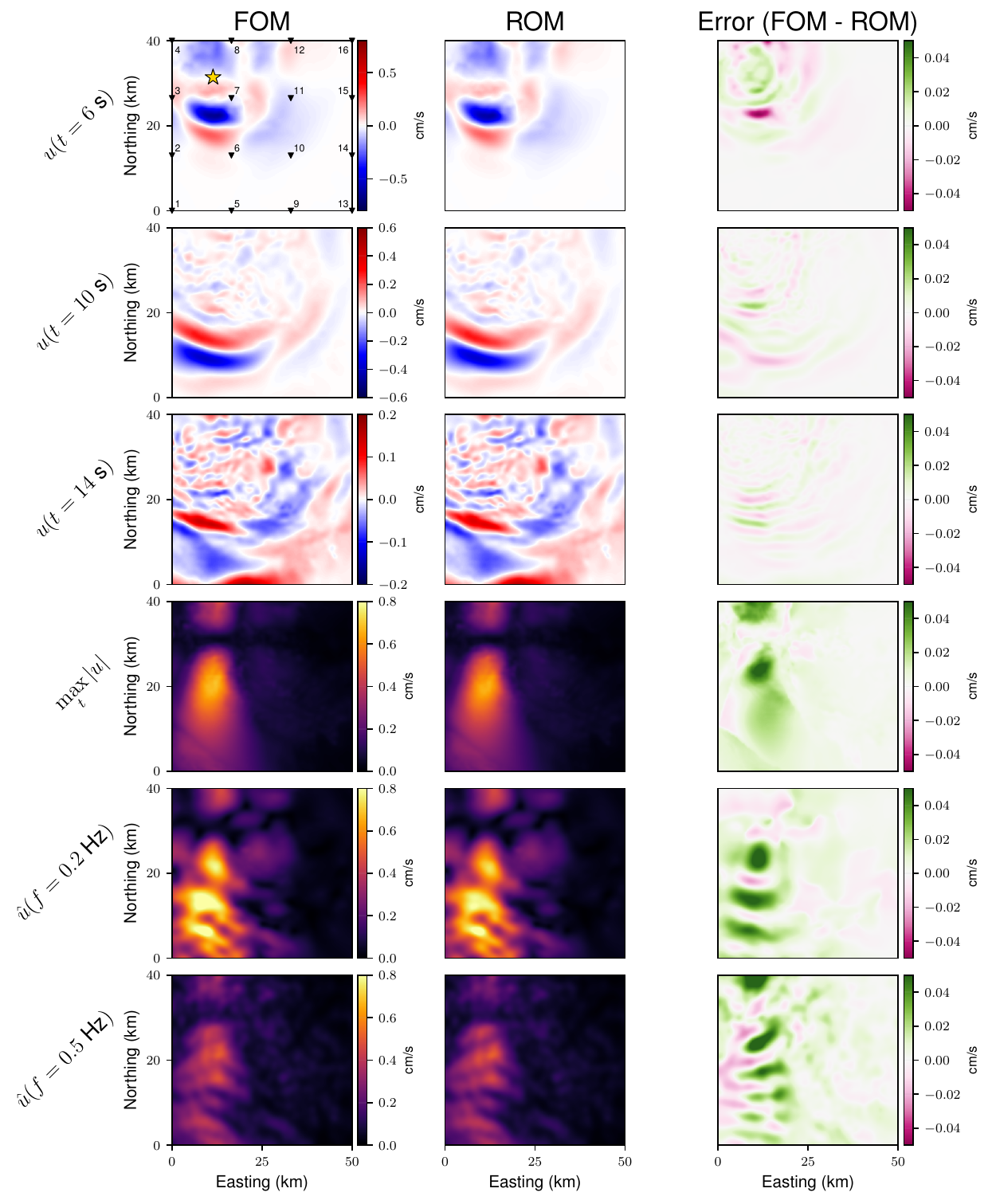}
    \caption{Surface wavefield snapshots and ground motion intensity maps,
    comparing the FOM results (left column), ROM results (middle column),
    and the difference between them (FOM - ROM, right column). The east-component ($u$)
    of the velocity wavefields are shown in the first three rows. The fourth row shows
    the maximum absolute velocity over all time steps, and the fifth and sixth rows show the
    Fourier amplitudes at 0.2 and 0.5 Hz. The source is an $M_W=5.0$
    earthquake with the same focal mechanism as the $\widehat{\bm M}_1$ elementary tensor.
    The source location is indicated by the gold star in the top left panel,
    and has a depth of 12.6 km.
    The locations of sixteen receivers used in subsequent plots are indicated
    by the black triangles in the top left panel. An animated movie
    of the FOM and ROM velocity wavefields showing all time steps is provided in the supplementary
    Movie S1.}
    \label{fig:rom_fom_m0}
\end{figure}

Fig. \ref{fig:rom_fom_m0} shows map-view comparisons of the ROM and FOM
surface wavefields at different timesteps of the simulation, as well as ground motion intensity
maps. In these results, we scaled the waveform amplitudes to correspond to
an $M_W=5$ earthquake (i.e., we multiplied them by a factor of 50).
The FOM and ROM show good agreement
in the velocity wavefields across all timesteps.
 The timing and amplitudes of the seismic
wave arrivals in the FOM are well reproduced by the ROM.
For this particular source location, a pronounced velocity pulse
travels southward from the epicenter, which results in the strongest ground shaking
in that direction. This amplification is likely influenced by the Los Angeles Basin
and local site effects that affect ground motions in the southwest.
Additionally, we observe significant scattering of the wavefield,
particularly apparent at 14~s simulation time, which is expected due to the uneven topography
and heterogeneity in the complex velocity model used.

Interestingly, the error (FOM - ROM) maps exhibit wave-like behavior, which is likely due
to the absolute amplitudes of the waves being slightly under-predicted.
Positive velocities (red) tend to correspond
to green areas (FOM $>$ ROM),
and negative velocities (blue) tend to correspond to pink areas (FOM $<$ ROM).
Overall, the ROM generally under-predicts the strength of ground motion
velocity as seen in the $\displaystyle\max_t|u|$ map. However, this under-prediction
reaches only about 5\% at worst.
Despite these minor inaccuracies, the FOM and ROM maintain good agreement
throughout the different geological features and various landscapes in our model domain.

\begin{figure}
    \centering
    \includegraphics[width=\textwidth]{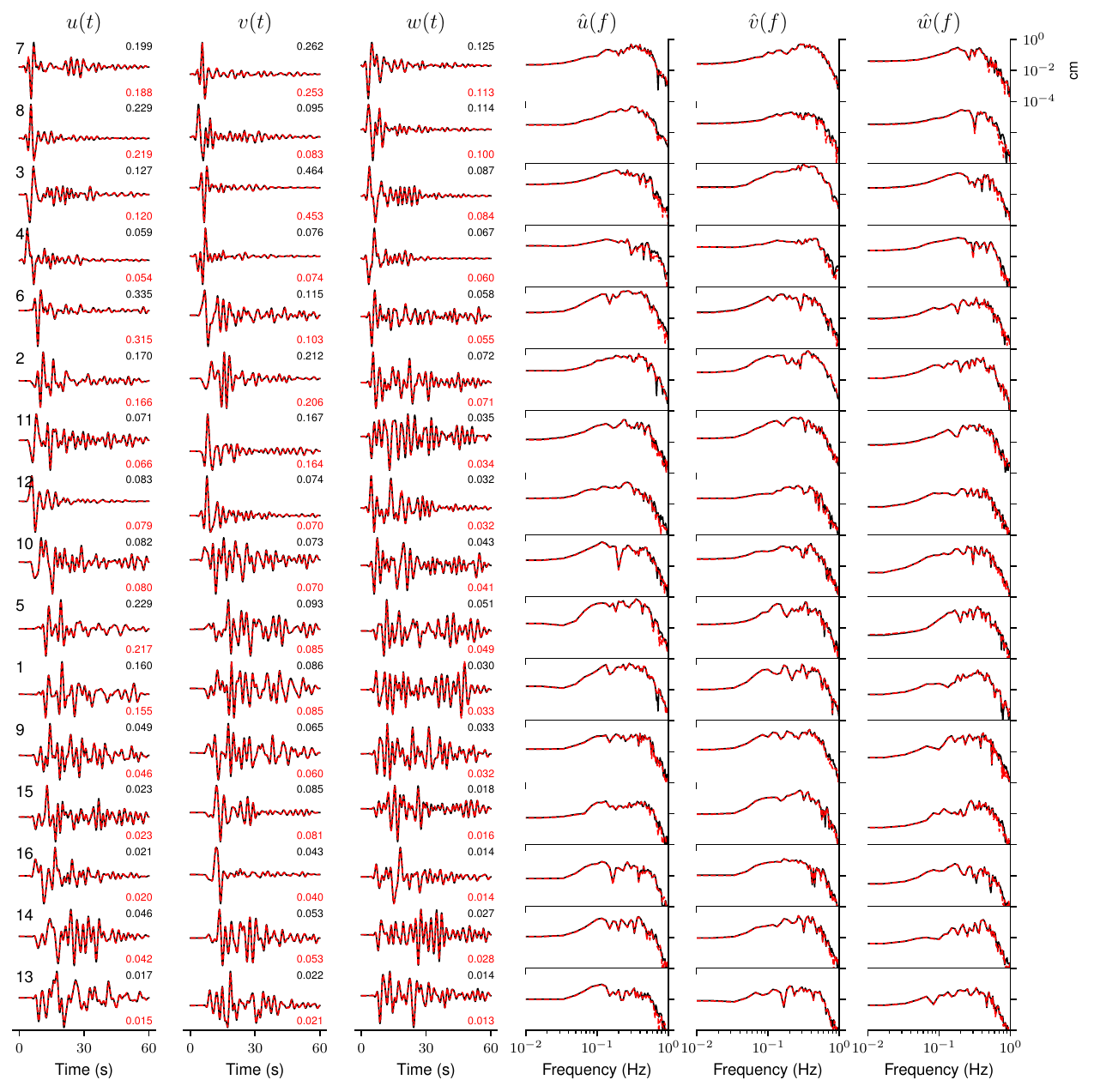}
    \caption{Three-component velocity waveforms
    ($u$, $v$, and $w$) and their Fourier transforms for the
    sixteen receivers labeled in Fig. \ref{fig:rom_fom_m0}.
    The earthquake source is the same as the source for Fig. \ref{fig:rom_fom_m0}.
    The FOM data are plotted with a solid black line,
    and the ROM data with a dashed red line. The receivers
    are ordered by increasing epicentral distance. The numbers on the right
    side of the waveform panels indicate maximum
    absolute value of the waveforms (FOM in black, ROM in red).}
    \label{fig:rom_fom_m0_w}
\end{figure}

The overall good fit of ROM ground
motion levels to the FOM data is also evident
in the Fourier amplitude spectra of the velocity seismograms (Fig. \ref{fig:rom_fom_m0_w}).
We computed the spectra by taking the Fourier transform of the velocity data,
taking the absolute value, and then multiplying by the sampling interval of 0.1 s.
We did not apply any smoothing to the spectra. The spectral amplitudes
at 0.2 Hz and 0.5 Hz show some slight under-prediction, as seen in the green areas in the right column of Fig. \ref{fig:rom_fom_m0}.
In all ground motion maps, stronger shaking occurs in the western
half of the simulation domain, which is also where ROM errors are larger.
Errors tend to be close to zero in the east, while
under-prediction in the west reaches 0.05 cm/s for $\displaystyle\max_t|u|$
at worst, and 0.05 cm for the spectra
at 0.2 and 0.5 Hz. At the higher frequency, the error is more spatially
variable and the ROM may alternate between over- and under-prediction
across short distances of less than 1 km.

By averaging across all of the $n_s$ discrete source locations,
we can compute the mean values for
the MAVE, MPGVE, and MSE metrics (as defined in Sec. \ref{sec:validation}) for each $j$ velocity component via
\begin{align}
    \overline{\textrm{MAVE}}_j&=\frac{1}{n_s}\sum_{i=1}^{n_s}\textrm{MAVE}_j(\bm{p}_i), \\
    \overline{\textrm{MPGVE}}_j&=\frac{1}{n_s}\sum_{i=1}^{n_s}\textrm{MPGVE}_j(\bm{p}_i), \\
    \overline{\textrm{MSE}(f)}_j&=\frac{1}{n_s}\sum_{i=1}^{n_s}\textrm{MSE}_j(\bm{p}_i, f).
\end{align}
The results are summarized in Table \ref{tab:err} for the four RBF kernels
that we tested (linear, cubic, quintic, and thin plate spline).
We tabulate the mean errors when the ROM-predicted data
are simply equal to the FOM data for the simulation
using the nearest-neighboring source location (``nearest'' column).

As shown, the cubic kernel gives the lowest $\overline{\textrm{MAVE}}$ for all
three components. When averaging over sources, receivers,
and time steps, the $\overline{\textrm{MAVE}}$ is about 0.007 cm/s for the two
horizontal components and about 0.004 cm/s for the
vertical component. The cubic kernel also gives the lowest
$\overline{\textrm{MSE}}$ when evaluated at 0.5 Hz. The quintic kernel performs
the best for the $\overline{\textrm{MPGVE}}$
metric and the $\overline{\textrm{MSE}}$ metric for 0.2 Hz. We discuss
these results further in Sec.~\ref{sec:discussion}, comparing
our results against empirical GMMs and other ML approaches.

\begin{table}[h]
    \centering
    \begin{filecontents*}{tables/errors.csv}
,nearest,linear,cubic,quintic,thin plate spline
$\overline{\textrm{MAVE}}_1$,0.0135,0.0085,\textbf{0.0069},0.0073,0.0072
$\overline{\textrm{MAVE}}_2$,0.0128,0.0081,\textbf{0.0067},0.007,0.0069
$\overline{\textrm{MAVE}}_3$,0.0079,0.0051,\textbf{0.0043},0.0046,0.0044
$\overline{\textrm{MPGVE}}_1$,0.0471,0.0516,0.0303,\textbf{0.0278},0.0359
$\overline{\textrm{MPGVE}}_2$,0.0446,0.0477,0.0286,\textbf{0.0265},0.0335
$\overline{\textrm{MPGVE}}_3$,0.0235,0.0203,0.0142,\textbf{0.0141},0.0157
$\overline{\textrm{MSE}}(f=0.2 \textrm{ Hz})_1$,0.0985,0.0687,0.0503,\textbf{0.0454},0.056
$\overline{\textrm{MSE}}(f=0.2 \textrm{ Hz})_2$,0.0964,0.0676,0.0496,\textbf{0.0444},0.0554
$\overline{\textrm{MSE}}(f=0.2 \textrm{ Hz})_3$,0.06,0.0402,0.0279,\textbf{0.0244},0.0318
$\overline{\textrm{MSE}}(f=0.5 \textrm{ Hz})_1$,0.0595,0.0837,\textbf{0.0534},0.0552,0.0607
$\overline{\textrm{MSE}}(f=0.5 \textrm{ Hz})_2$,0.0559,0.0775,\textbf{0.0500},0.0522,0.0565
$\overline{\textrm{MSE}}(f=0.5 \textrm{ Hz})_3$,0.0309,0.0355,\textbf{0.0237},0.0239,0.0267
\end{filecontents*}
\csvautotabular{tables/errors.csv}
    \caption{Summary of the mean absolute velocity error
    (MAVE), mean peak ground velocity error (MPGVE),
    and mean spectral error (MSE) obtained from leave-one-out
    cross-validation for different interpolators. The errors are reported in
    cm/s for MAVE and MPGVE, and in cm for MSE. The
    velocity waveform amplitudes have been scaled to correspond
    to an $M_W$ 5.0 earthquake.
    The error measurements indicate the averages for all source locations, all receivers,
    and for all elementary moment tensors in the simulated dataset.
    The lowest error for each metric is indicated in bold font.}
    \label{tab:err}
\end{table}

\begin{figure}
    \centering
    \includegraphics[width=\textwidth]{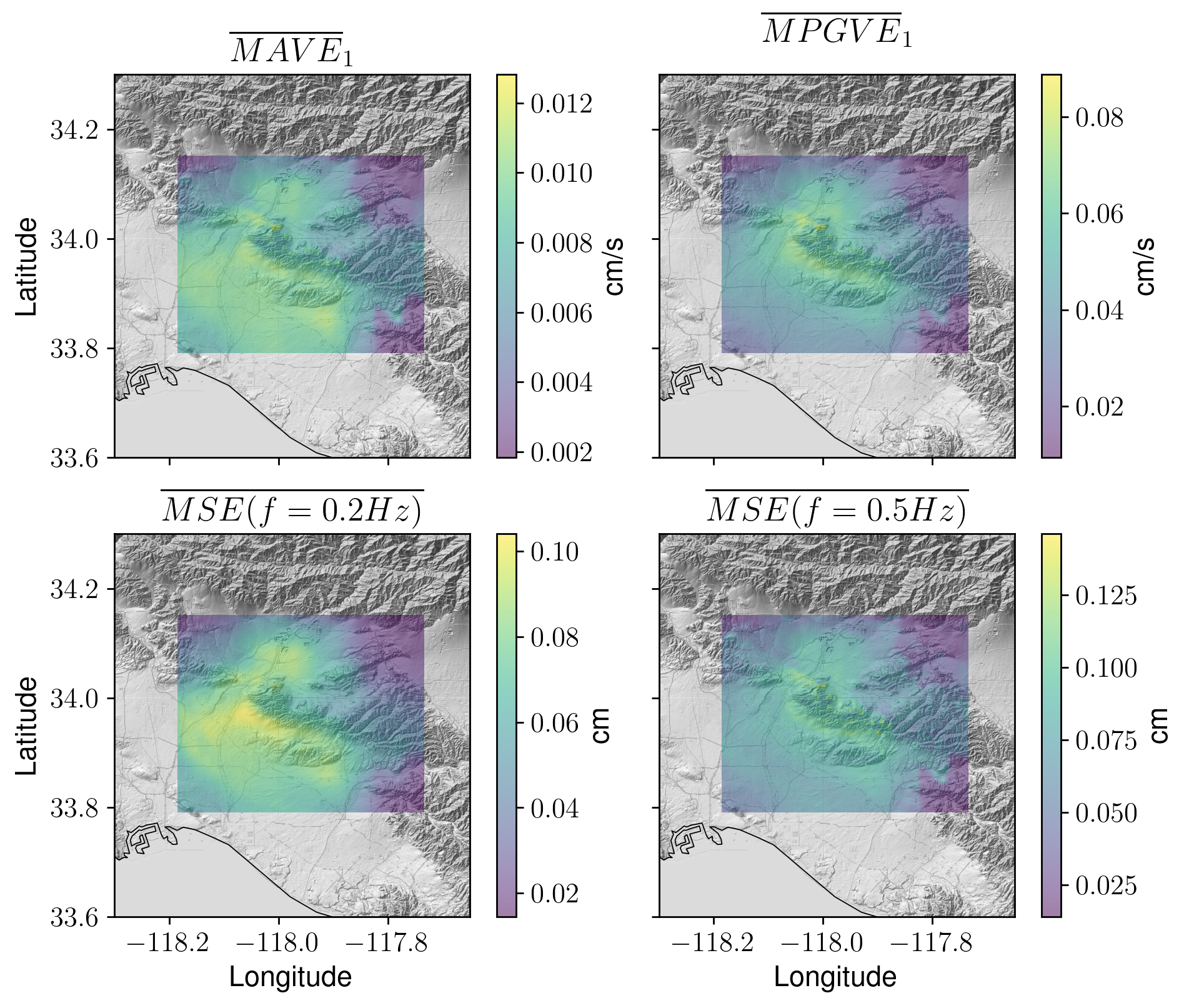}
    \caption{Map views of the leave-one-out ROM
    mean absolute velocity error (MAVE), mean peak ground velocity error
    (MPGVE), and mean spectral error (MSE). The errors are averaged
    averaged over all elementary moment tensor sources in the dataset,
    and are shown for the east ($u$) component of the
    surface wavefield. The errors are plotted on top of the topography
    in southern California and the amplitudes
    correspond to an $M_W=5.0$ earthquake.}
    \label{fig:rom_error}
\end{figure}

We examine the spatial trends in the source-averaged
errors in Fig. \ref{fig:rom_error}. For all metrics,
the errors tend to be largest closer to the center of the domain
and smallest in the furthest corners. One exception is the southwest
corner where the Los Angeles basin is likely amplifying ground
motion. Interestingly, the error also correlates with some topographic features.
This could be due to a correlation
between topography and seismic wave speeds near the surface,
or due to topographic amplification and shielding effects \citep[e.g.,][]{chaljub2010quantitative,hartzell2014ground}.
For example, receivers in the Los Angeles Basin and San Gabriel Valley
show higher errors than receivers located in the Puente Hills and other
foothills of the Santa Ana and San Gabriel Mountains (Fig. \ref{fig:rom_error}).

When examining the seismograms that are recorded by individual receivers,
we can inspect the amplitudes and timing of the seismic waves
directly. We picked out sixteen evenly spaced receivers at the surface of the
domain and plotted the velocity time series for the three components of the wavefield (Fig. \ref{fig:rom_fom_m0_w}).
The locations of the sixteen receivers are indicated in the top left
panel of Fig. \ref{fig:rom_fom_m0}. We also computed the
Fourier transforms of the velocity timeseries for each component and plot
the spectral amplitudes as
a function of frequency ($\hat{u}(f)$, $\hat{v}(f)$, and $\hat{w}(f)$)
from 0.01 Hz to 1.0 Hz.
Note that the time series are lowpass filtered
at 0.5 Hz which explains the rapid decay in spectral amplitudes starting
around that frequency.

The seismograms and Fourier spectra show good agreement,
with the FOM and ROM curves almost always plotting
nearly exactly on top of each other. All of the wiggles in the
FOM velocity time series are also present in the ROM time series, showing
that the ROM reproduces the entire wavefield and accurately
captures the duration of ground shaking.
The Fourier spectra also show how the amplitudes of the
spectral content is very similar between the FOM and ROM. The
spectra for some receivers
show troughs or peaks in spectral amplitude at narrow frequency bands,
and these are also accurately reproduced by the ROM.

\subsection{ROM ground motion estimates for general point source
moment tensor sources}

In addition to the six elementary moment tensors ($\widehat{\bm M}_1$ through $\widehat{\bm M}_6$),
the ROM and FOM can be used to produce waveforms for general moment tensors
expressed as a combination of the six elementary tensors.
As a demonstrator, this section focuses on a particular seismic moment tensor given by
\begin{equation}
	\bar{\bm M} =
    \begin{bmatrix}
        0.56 & 1.87 & 2.63 \\
        1.87 & 3.11 & 1.69 \\
        2.63 & 1.69 & -3.67
    \end{bmatrix} \times 10^{14} \ \text{N}\, \text{m}
    \label{eq:mt}
\end{equation}
and located at ($\Delta\ell=11.2$ km, $\Delta w=4.9$ km, $\Delta z=12.0$ km)
which is not part of the training dataset.
We chose this moment tensor as it corresponds to the first subfault in a
finite fault rupture model that we analyze in Sec.~\ref{sec:ff}. Upon decomposition
into the \cite{kikuchi1991inversion} basis,
we find the following weights
\begin{align*}
    c_1&=1.87\,, \\
    c_2&=-3.11\,, \\
    c_3&=1.69\,, \\
    c_4&=2.63\,, \\
    c_5&=-3.67\,, \\
    c_6&=0\,.
\end{align*}
We can use these weights to construct the seismograms
for this moment tensor by plugging them into Eq. (\ref{eq:recon_mt}). Since the
ROM must be evaluated five times (not six, since
$c_6=0$), it takes about five times more computing
time to obtain the full wavefield solutions for this
seismic source, compared to the elementary moment tensor
sources in Sec.~\ref{sec:res-m0}.

\begin{figure}
    \centering
    \includegraphics[width=\textwidth]{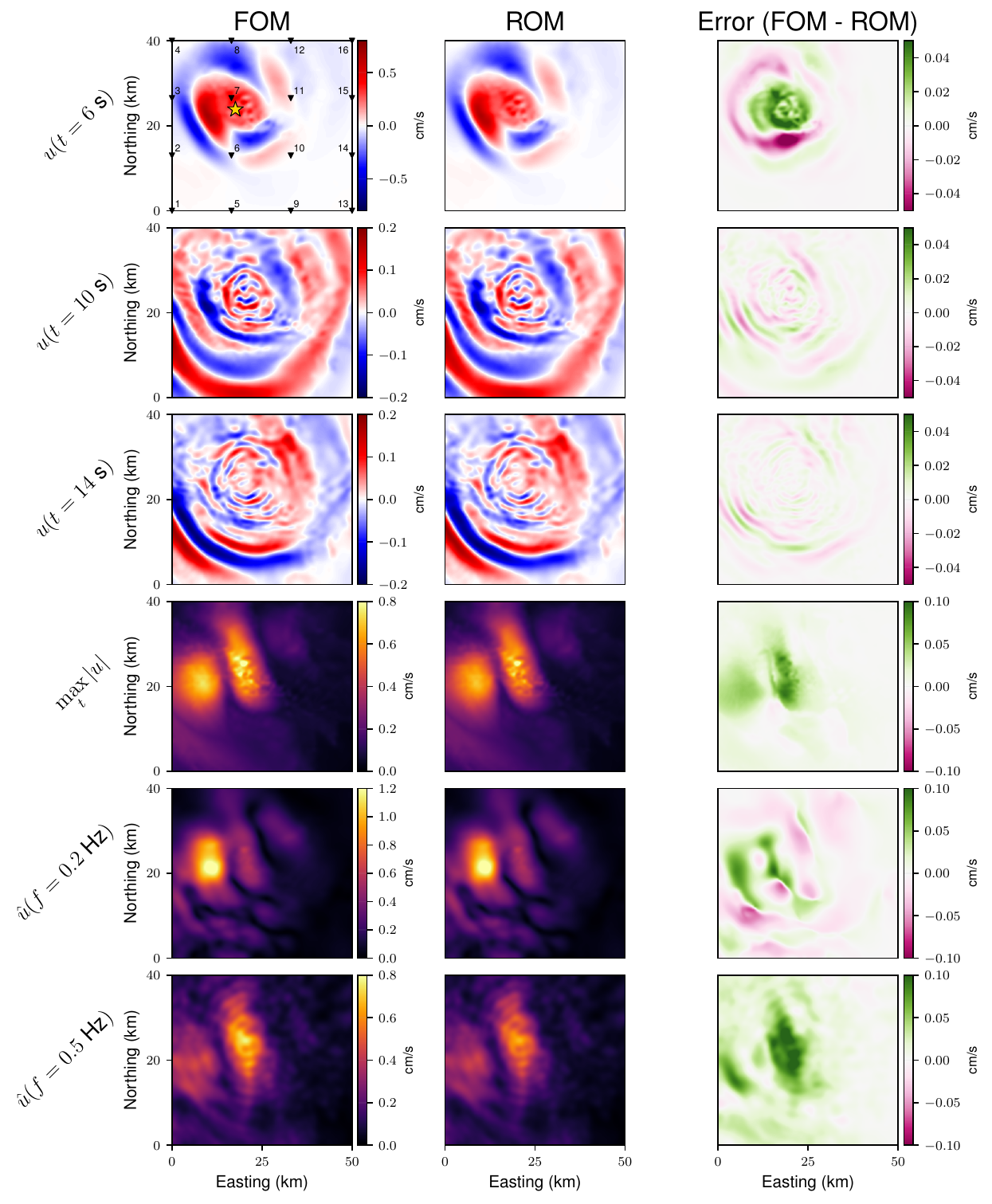}
    \caption{Wavefield snapshots and ground motion intensities,
    comparing the FOM against the ROM, for a general moment tensor.
    The figure format is the same as Fig. \ref{fig:rom_fom_m0}, but is for
    a different moment tensor source given in Eq. (\ref{eq:mt}).
    The source location is indicated by the gold star in the top left panel,
    has a depth of 12.0 km, and a moment magnitude of 5.0.
    An animated movie of the FOM and ROM velocity wavefields showing
    all time steps is provided in the supplementary Movie S2.}
    \label{fig:rom_fom_mt}
\end{figure}

\begin{figure}
    \centering
    \includegraphics[width=\textwidth]{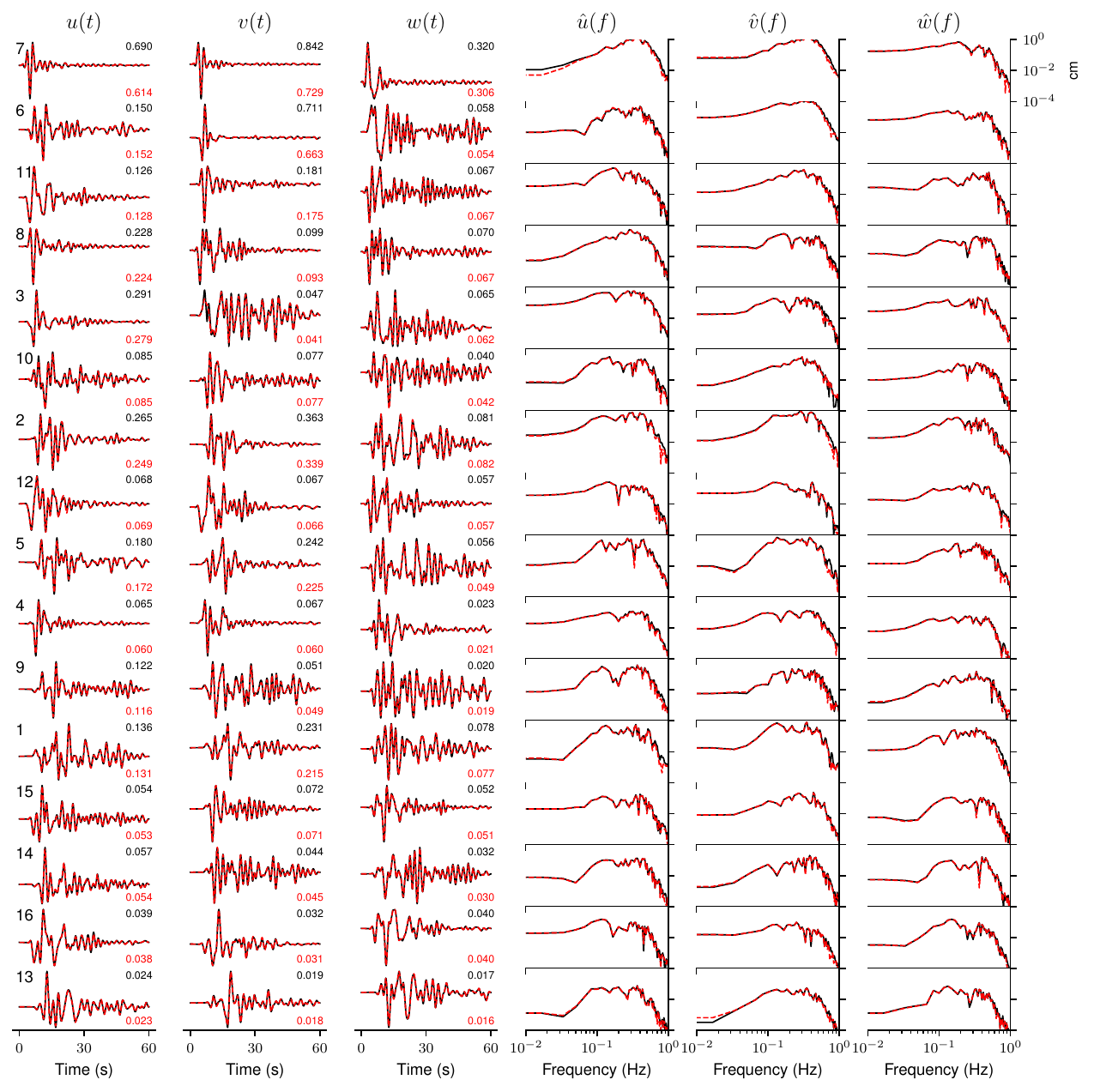}
    \caption{Three-component velocity waveforms
    ($u$, $v$, and $w$) and their Fourier transforms for the
    same moment tensor source shown in Figure \ref{fig:rom_fom_mt}.
    The FOM data are plotted with a solid black line,
    and the ROM data with a dashed red line. The numbers on the right
    side of the waveform panels indicate maximum
    absolute value of the waveforms (FOM in black, ROM in red).}
    \label{fig:rom_fom_mt_w}
\end{figure}

Similar to the elementary moment tensor source, we plot
the map-views of the seismic wavefields (Fig. \ref{fig:rom_fom_mt}) and the
velocity timeseries and Fourier spectra recorded at the same sixteen receivers
(Fig. \ref{fig:rom_fom_mt_w}).
Note that this source is at a slightly different location
and depth than the source for Figs. \ref{fig:rom_fom_m0}
and \ref{fig:rom_fom_m0_w}. Despite this change in source,
the FOM and ROM wavefields still match well, with differences
in velocity less than 0.05 cm/s. Similar
to the $\widehat{\bm M}_1$ source results, the maximum absolute velocities tend
to be slightly under-predicted at worst by about 0.05 cm/s in the epicentral region.
The higher frequency 0.5 Hz spectra are also slightly under-predicted
in the same area.

Despite small discrepancies, these results
instill confidence in our procedure for combining the solutions for
different elementary moment tensors. They demonstrate that the
errors are not significantly larger for general moment tensors compared
to the elementary moment tensors.

\subsection{ROM ground motion estimates for finite sources}
\label{sec:ff}

For moderate to large magnitude earthquakes, the point
source approximation is often inaccurate for predicting
ground motions. Larger earthquakes are typically
better modeled as slip distributed on a finite fault plane \cite[e.g.,][]{IDE2007193,hayes2017finite}.
These types of earthquake rupture models can also account for the
rupture speed and direction of rupture propagation by
modeling the source time function for a set of discrete points
on the fault. Accounting for the finite source effects such as rupture directivity can strongly influence
ground motion and cause strong asymmetry of the ground
motion amplitudes \cite[e.g.,][]{spudich2008directivity}.

As a demonstration example of using our ROMs to predict ground motions for
larger magnitude earthquakes, we simulate seismic ground motions for a published kinematic
rupture model of the 1987 $M_W$ 5.9 Whittier Narrows
earthquake. This blind thrust earthquake is now believed to have occurred
as part of the Puente Hills thrust system \citep{shaw2002puente}
which is situated under
the Los Angeles metropolitan area and poses a significant risk to the region \citep{field2005loss}.
\cite{hartzellSourceComplexity19871990} modeled the earthquake
using a kinematic source description, and we use their information on model parameterization that is available from the SRCMOD database \citep{maiSRCMODOnlineDatabase2014}.
Their rupture model is parameterized by 100 subfaults that create a square fault
with a side length of 10 km. The fault is modeled using a strike of 280
degrees and a dip of 30 degrees. See Fig. \ref{fig:map} for the location
of the finite fault in our modeling area.
The total slip amplitudes are available from the SRCMOD database,
but not the variable dip-slip and strike-slip components.
We calculate the dip-slip and strike-slip components
by assuming a uniform rake angle of 58$^{\circ}$ following the
focal mechanism solution produced by the Southern
California Seismic Network \citep{hutton2010earthquake}.

\begin{figure}
    \centering
    \includegraphics[width=\textwidth]{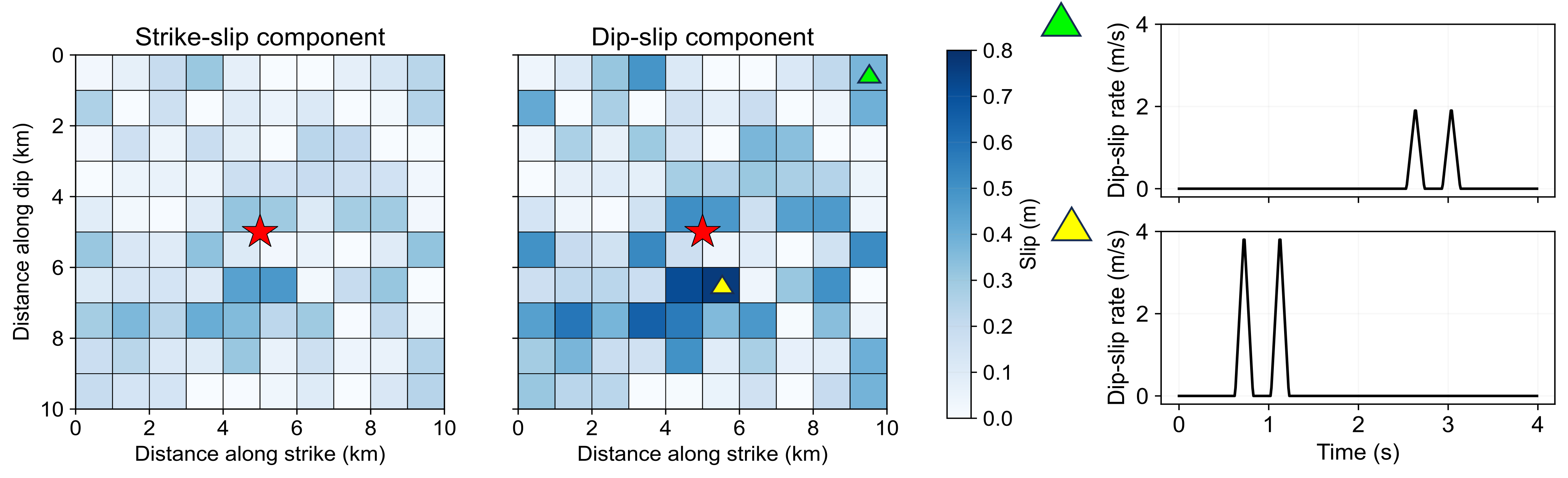}
    \caption{Finite fault model for the 1987 $M_W$ 5.9 Whittier Narrows
    earthquake described in \cite{hartzellSourceComplexity19871990}. The
    images show the strike-slip component, dip-slip component, and
    slip rate time histories for two of the subfaults in the model.}
    \label{fig:ff}
\end{figure}

\cite{hartzellSourceComplexity19871990} developed
a few different slip models for this earthquake, using the same parameterization.  We specifically use their ``L18'' model
which is available from the SRCMOD database \citep{maiSRCMODOnlineDatabase2014}.
The slip distribution for this model is plotted in Fig. \ref{fig:ff}.
This model assumes a constant rupture velocity, starting from the hypocenter, and allows
each subfault to rupture twice. The source-time function for each
subfault contains two triangular pulses, each with a 0.2 s duration
and a time separation of 0.2 s (Fig. \ref{fig:ff}). We converted this kinematic
rupture model into the Standard Rupture Format \citep{graves2014srf}
and then performed a kinematic SeisSol simulation to obtain the
seismograms using the FOM procedure described in Sec.~\ref{sec:ff}.

To obtain the ROM seismograms, we use the
approximate Green's function and discrete representation theorem,
using Eqs. (\ref{eq:approx_green}) and (\ref{eq:recon_ff}).
For this simulation, we must obtain the ROM seismograms
using the representation theorem
approach as the source time function used by
\cite{hartzellSourceComplexity19871990} is different from the source time
function, $\dot{M}(t)=\frac{M_0t}{T^2}$, that
we used at the 500 discrete training locations.

\begin{figure}
    \centering
    \includegraphics[width=\textwidth]{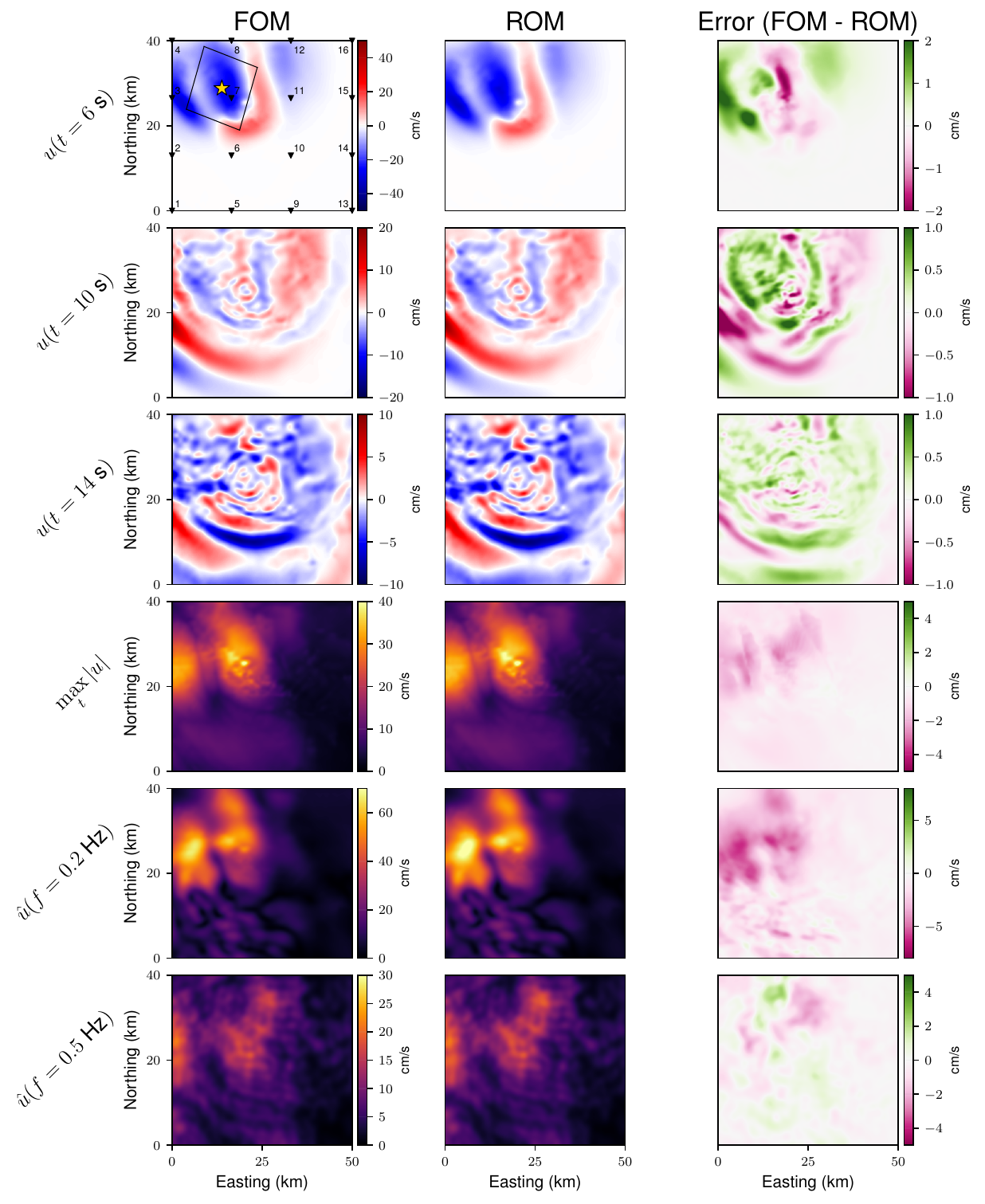}
    \caption{Wavefield snapshots and ground motion intensities,
    comparing the FOM against the ROM for the
    1987 $M_W$ 5.9 Whittier Narrows finite fault simulation.
    The source model is the ``L18'' kinematic rupture model
    from \cite{hartzellSourceComplexity19871990}.
    The rupture extent of the finite fault is outlined by the black
    rectangle in the top left panel.
    An animated movie of the FOM and ROM velocity wavefields showing
    all time steps is provided in the supplementary Movie S3.}
    \label{fig:rom_fom_ff}
\end{figure}

The FOM and ROM wavefields match well for
this finite fault rupture simulation (Fig. \ref{fig:rom_fom_ff}). The error
between the FOM and the ROM velocity amplitudes tends to be no larger than
5\% across all time-steps in the simulation.
This result is comparable to the results for the general point source
moment tensor where the error in the velocity wavefields was
also approximately 5\% at worst. This result is somewhat surprising given that
the finite source uses a significantly more complex rupture model that has
many point sources each with a different source time function.

\begin{figure}
    \centering
    \includegraphics[width=\textwidth]{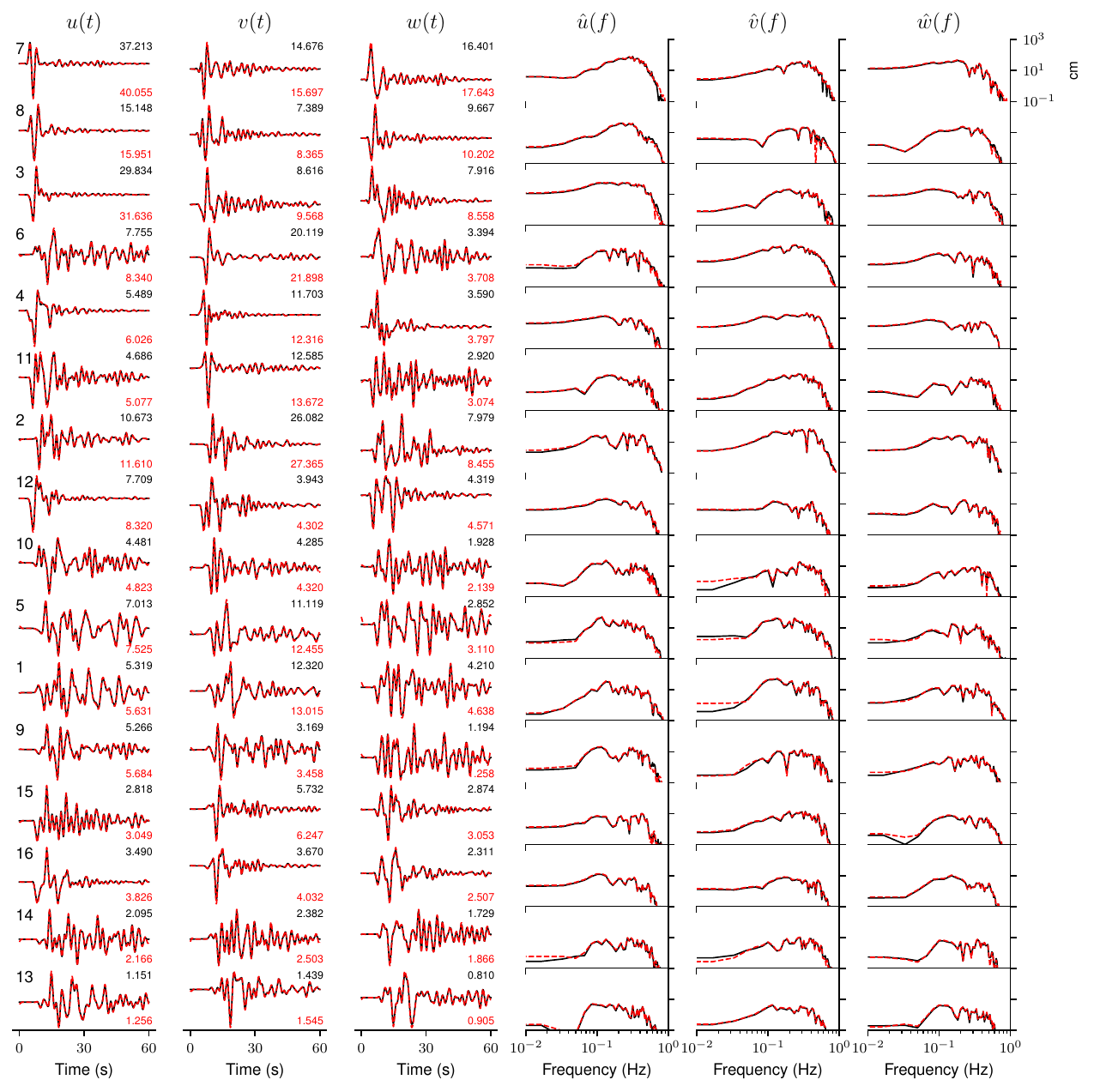}
    \caption{Three-component velocity waveforms
    ($u$, $v$, and $w$) and their Fourier transforms for the
    Whittier Narrows finite fault source.
    The FOM data are plotted with a solid black line,
    and the ROM data with a dashed red line. The numbers on the right
    side of the waveform panels indicate maximum
    absolute value of the waveforms (FOM in black, ROM in red).}
    \label{fig:rom_fom_ff_w}
\end{figure}

In our finite fault model results, the ROM tends to
slightly over-predict the ground motion amplitudes as seen
by the mostly pink colors in the error panel for $\displaystyle\max_t|u|$.
This might be related to the low-frequency Fourier spectra being over-predicted
by the ROM, as the 0.2 Hz panel shows only over-prediction, while the
0.5 Hz panel shows both over- and under-prediction. In the velocity seismograms
and Fourier spectra (Fig. \ref{fig:rom_fom_ff_w}), the FOM and ROM also
show good agreement. As noticed in the wavefields, the ROM absolute maximum
velocities tend to be slightly larger than the FOM values. This over-prediction
is only at worst about a few cm/s, as seen in receiver number 7 located
directly above the fault where the over-prediction is about 3 cm/s.

Obtaining the ROM seismograms for the finite fault simulation requires significantly
more computational resources than the point source simulations. The cost of the ROM
solutions for finite fault sources scales approximately linearly with the number of
subfaults, if we assume one general moment tensor source representing each subfault in the FOM,
as the finite fault runtime (0.71 CPUh) is about 100 times the
runtime of the general moment tensor source (0.0067 CPUh).
This finite fault runtime also includes the time spent computing the Green's
functions, multiplication in the frequency domain, and computing the
inverse Fourier transforms, which explains why the scaling is not
perfectly linear. Despite this additional cost, the ROM is still two orders
magnitude faster than the FOM. However, if more than 100
moment tensor point sources are used to represent the kinematic model,
then the computational speedup achieved by using the ROM would be smaller.
See Table \ref{tab:runtime} for the full list of runtimes
and speedups for the FOM and ROM for different source
descriptions.

\begin{table}[h]
    \centering
    \begin{filecontents*}{tables/runtimes.csv}
,FOM (CPUh),ROM (CPUh),Speedup
Single elementary tensor,192,0.0011,1.75E+05
General moment tensor (6 components),192,0.0067,2.87E+04
Finite fault (100 sub-faults),198,0.7101,2.79E+02
\end{filecontents*}
\csvautotabular{tables/runtimes.csv}
    \caption{FOM and ROM runtimes. The numbers indicate the CPUh
    usage required to generate the complete 60-second,
    three-component velocity waveforms
    for all 8,181 receivers. The ROM value for the
    single elementary tensor indicates
    the time spent performing the summation of the POD basis
    using Eq. (\ref{eq:recon_elem}). The general moment tensor row
    indicates the time spent solving for the elementary moment tensor
    weights in Eq. (\ref{eq:elem_weights}) and summing the six-component
    POD basis (Eq. (\ref{eq:recon_mt})).
    The ROM time for the finite fault includes the time to calculate
    elementary weights in Eq. (\ref{eq:elem_weights}),
    compute seismograms using Eq. (\ref{eq:recon_mt}),
    compute approximate Green's function using Eq. (\ref{eq:approx_green}),
    and perform the convolution and summation in Eq. (\ref{eq:recon_ff}).
    The speedup is calculated by dividing the FOM column by the ROM column.
    The time spent computing the SVD is not included in the ROM
    runtimes. We measured the SVD time to be approximately 87 seconds
    on a single core on SuperMUC-NG using an Intel Skylake
    Xeon Platinum 8174 processor and a single thread. The SVD time decreased to
    about $\approx$20 seconds when using 16 threads.}
    \label{tab:runtime}
\end{table}

\subsection{Real data comparison for a Los Angeles area earthquake}
\begin{figure}[H]
    \centering
    \includegraphics[width=\textwidth]{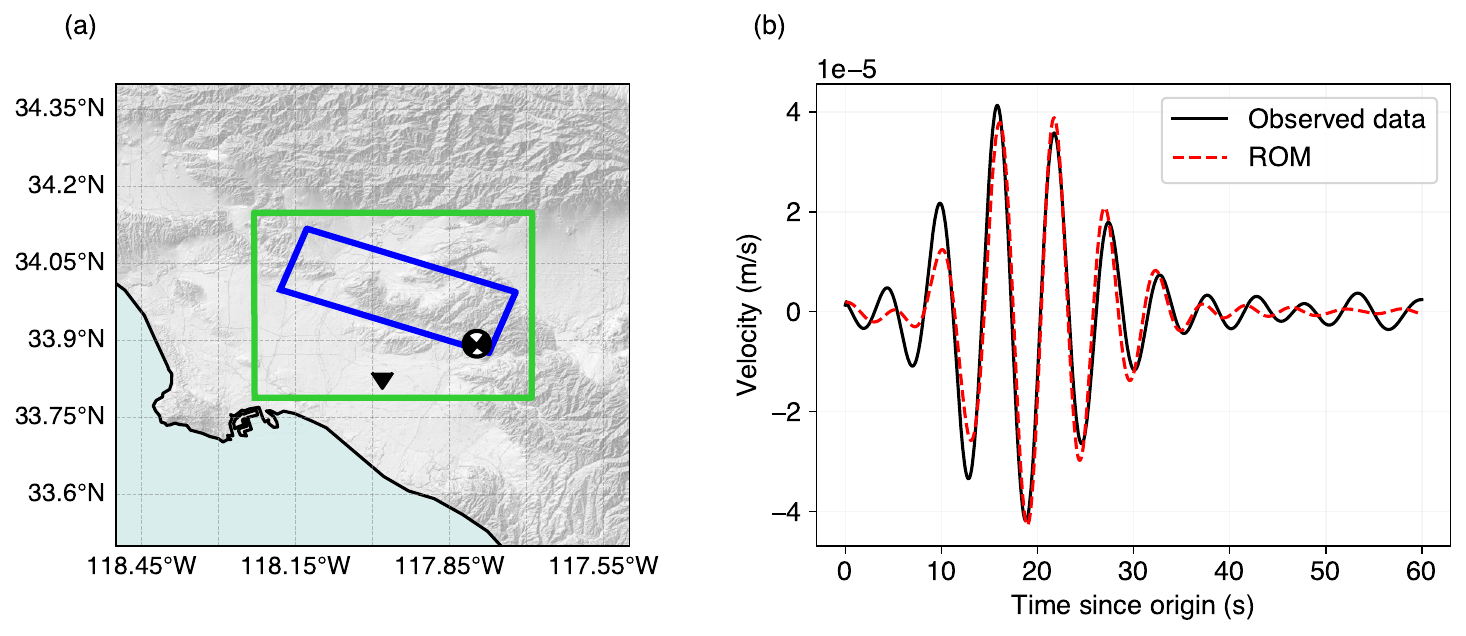}
    \caption{(a) Location of a real earthquake (USGS event ID ci10399801)
    that occurred in 2009 and the station CI.BRE we used to validate the ROM with real data.
    The earthquake moment magnitude is 3.55 and the depth is 5.0 km.
    We use the moment tensor solution produced by the Southern
    California Seismic Network \citep{hutton2010earthquake}. (b) We plot the tangential component
    of the velocity, in solid black, using the data from the broadband seismometer at
    station CI.BRE. We processed the data by removing the mean,
    removing the instrument response, bandpass filtering from 5-10 seconds
    following \cite{lai2020shallow},
    and rotating to obtain the tangential component.
    We also plot the filtered ROM velocity data for the tangential
    component, in dashed red, at the nearest receiver for this
    moment tensor solution.}
    \label{fig:real_seismo}
\end{figure}

To demonstrate using our ROM for simulating real earthquakes,
we selected an earthquake in the Los Angeles area with
USGS event ID ci10399801. This was an $M_W$ 3.55 earthquake that occurred in
2009 with a depth of 5.0 km and was located towards near the southeast corner
of our source region (Fig.~\ref{fig:real_seismo}).
This earthquake was included in the simulations performed by \cite{lai2020shallow}
who also used CVM‐S4.26.M01 and found good agreement between
real and simulated data on the tangential
velocity component at station CI.BRE, the Southern California
Edison Barre Peaker power station in Stanton, CA. This recording
exhibits strong amplification due to the Los Angeles Basin.
We obtained the recorded velocity data from the broadband seismometer
at station CI.BRE from the Southern California Earthquake Data Center
\citep{center2013southern} and processed the data by removing the mean,
removing the response, bandpass filtering from 5-10 seconds,
and rotating to the tangential component.
To obtain the ROM velocity data, we used the Southern California Seismic Network
moment tensor solution \citep{hutton2010earthquake}
and Eq. (\ref{eq:recon_mt}) to obtain the predicted velocity data.
We then selected the receiver that was located nearest to station CI.BRE.
The bandpass filtered velocity data show good agreement, as the timing and amplitudes
on the tangential component match well between the ROM and observed data (Fig.~\ref{fig:real_seismo}).

\section{Discussion}
\label{sec:discussion}
\subsection{Comparison with Previous Work}
This work demonstrates that simulated seismic wavefields can be accurately
approximated using a non-intrusive reduced-order model to obtain rapid solutions for earthquakes described by moment
tensor point sources or finite slip models in a region of southern California.
This work significantly extends the ROMs presented in \cite{rekoske2023}
which generate PGVs, rather than complete waveforms, for only a single epicentral location.
We show that our time-dependent approach is applicable to finite sources
by exploiting superposition of the wave equation.

The field of using machine learning or other interpolation approaches for simulating seismic wave propagation is rapidly advancing \citep[e.g.,][]{yang2021seismic,yang2023rapid,lehmann20243d}.
However, direct one-to-one comparison with our work is challenging, especially in terms of accuracy and computational cost.
Several methods have demonstrated rapid seismic waveform
modeling in 2D \citep{moseley2020deep,rasht2022physics,yang2021seismic,yang2023rapid},
but rapid 3D seismic wave propagation is more challenging.
Some of the recently proposed approaches using 3D wave propagation simulations
develop methods for complete 3D wavefields \citep{kong2023feasibility,zou2023deep}
while others predict 2D wavefields at the surface \citep{lehmann20243d,li2023rcnn}.
\cite{lehmann20243d} used a
factorized Fourier neural operator to simulate surface wavefields for
a family of 3D velocity models and reported a relative RMSE of 17\% for point sources.
Due to the many choices involved in simulating seismic wave propagation, it is difficult to directly
compare results head-to-head. Some of these choices include: (1)
the resolved frequency band, (2) the constitutive law (i.e., acoustic,
elastic, or visco-elastic), (3) the dimension (i.e., 2D or 3D) and size
of the simulation domain, (4) the spatial heterogeneities due to
velocity model, attenuation, and topography, and (5) the spatial
and temporal resolution of the surface wavefields.
Comparing with \cite{lehmann20243d}, we note that the maximum frequency
resolved in our seismograms is lower (e.g., 0.5 Hz compared to 5 Hz),
though our wavefield area is larger (2000 km$^2$ compared to $\sim$100 km$^2$),
the duration of our seismograms is longer (60 s compared to 7 s), and we
include more receivers (8181 compared to 256). Despite the lower frequency
we resolve, our FOMs are more computationally expensive ($\approx$200 CPUh
vs. $\approx$20 CPUh) as we model a larger area and slower wavespeeds (500
m/s vs. $\sim$1000 m/s) and account for visco-elastic attenuation,
which adds $\approx$80\% computational cost when using SeisSol \citep{uphoff2016generating} but may increase computational cost differently using other implementations.
Given that our realistic model setup includes
topography and slow wavespeeds, and our small average error of 0.007 cm/s for an
$M_W$ 5.0 earthquake, we think that our ROM approach offers an advantage
when a preferred, single velocity model
is chosen for wavefield prediction; neural operators may offer better generalizability when
using families of velocity models or performing seismic tomography.

\begin{figure}[H]
    \centering
    \includegraphics[width=\textwidth]{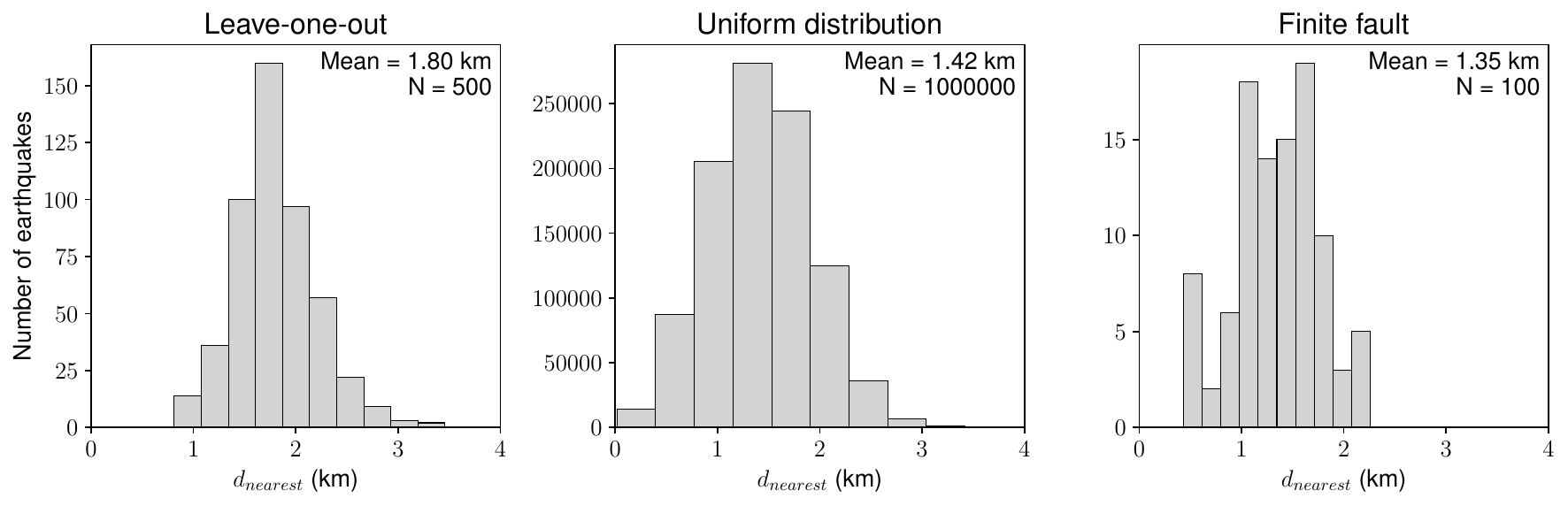}
    \caption{Histograms showing the distribution of $d_{nearest}$
    for three different datasets.
    The left panel shows the histogram for the Halton sequence
    used for performing the leave-one-out cross validation to
    determine the ROM accuracy. The middle panel shows the distribution
    for a uniform distribution in the parameter space that contains
    $10^6$ source locations. The right panel shows the distribution
    for the source locations in the Whittier Narrows finite fault
    model.}
    \label{fig:hists}
\end{figure}

One advantage of using RBFs for interpolating the POD coefficients in our work is that
LOOCV errors can be computed at very little computational expense. When using
large neural networks, computing LOOCV errors would typically require an expensive
retraining procedure, such that errors are usually reported for only one
train/test split instance. The LOOCV errors we present here are conservative
estimates of the true ROM error, i.e., the errors in real-time application of
the ROM for new unseen earthquakes would likely be lower due to the smaller
average spacing between sources (Fig.~\ref{fig:hists}). For any one of the training source
locations, the average distance to the nearest neighboring source location
$d_{nearest}$ is 1.80 km; for any earthquake that might occur in the entire
volume described by $\mathcal{P}$ the average $d_{nearest}$ is slightly smaller at 1.42 km
(Fig.~\ref{fig:hists}). In contrast to LOOCV, relying on a single test/train split
provides just one estimate of model performance, and the evaluation
might not reflect the true performance of the model on unseen data.

Our study also provides a new perspective on the efficient use of numerically calculated
elastodynamic Green's functions. Previous work using interpolated Green's functions
is also typically based on reciprocal simulations \cite[e.g.,][]{wang2020moving}.
Others have suggested that for applications using multi-linear interpolation of numerically precomputed Green's functions, their spacing
should be no greater than a quarter of the minimum wavelength \citep{heimann2019python}.
The minimum wavelength is calculated by dividing the minimum wave velocity
by the maximum frequency. In our 0.5 Hz accurate FOM forward simulations with $V_{S,min}=500$ m/s, the minimum wavelength is 1 km,
which would require a minimum spacing of Green's functions of 250 m for accurate multilinear
interpolation. Our results (e.g., Figs. \ref{fig:rom_fom_m0} and \ref{fig:rom_fom_m0_w}) suggest that a Green's function spacing
even as large as about 2 km, which is nearly twice the minimum wavelength, is sufficient
to produce accurate ROM interpolations when using interpolated POD with RBFs.
Other work suggests that it is superior to interpolate Green's functions using the
frequency-wavenumber domain \citep{gulunay2003seismic}, though a thorough comparison between
interpolation methods in various domains is outside the scope of this paper.
Future work may investigate other spectral decompositions, manifold interpolation,
or shifted proper orthogonal decomposition to better decompose advection-dominated
wave equations \citep[e.g.,][]{reiss2018shifted,lu2020lagrangian}.

\subsection{Limitations}
One limitation in our study relates to limited resolution we chose for the forward
simulations, as the maximum frequency content in our seismograms of 0.5 Hz is lower than
required for most engineering applications. We do not interpolate the waveforms in time,
meaning that the ROM always generates 60-second waveforms sampled at 10 Hz. However, this
is not a disadvantage of the ROM approach but merely a choice we made to limit our simulation
computational costs; higher resolution forward simulations would readily allow extending
our ROM approach to higher frequencies. A true disadvantage of this ROM approach is that
predicting the seismic wavefields for a different velocity model is not easily possible.
For example, several alternative 3D velocity models are available for this region that we
could have used, such as CVM-H or CVM-S, and it would be interesting to apply our ROMs to
different regions such as northern California or Cascadia. However, with the current approach,
as many new, expensive simulations would be required for each velocity model.

We note that the complex issue of nonlinearities in ground motion prediction
is a substantial consideration, especially for large earthquakes
\citep[e.g.,][]{field1997nonlinear,rubinstein2004evidence,roten2014expected}.
At this time, our FOM simulations do not account for soil nonlinearity or
off-fault plasticity which may influence the strength of ground motions.
Our approach for simulating large earthquakes
by the summation of point sources would not be applicable
for nonlinear wavefield effects, as the linear superposition of seismic sources
would be broken in the nonlinear case. Additional work or methods are needed to consider
how to rapidly obtain physics-based simulation results that would consider
nonlinear source or wavefield effects for which the principle of superposition
of the wave equation is no longer valid.

\subsection{Future work and implications}
One advantage of predicting complete timeseries data instead of individual ground motion
metrics is that any ground motion metrics can be computed from the
timeseries data. We focused on a few select metrics, such as the MPGVE
and MSE, that could be relevant for earthquake early warning and seismic
hazard applications. However, if new ground motion metrics become standard
in the engineering community, then those can be computed using the same
FOM data. Another improvement in the present study is that we allow for
varying epicentral locations and moment tensors. This feature allows
one to compute accurate wavefields for sources distributed throughout
the whole volume, not just for a set of assumed faults. This is
especially important in an area such as the Los Angeles Basin
which is known for having blind thrust earthquakes on faults
that were only identified after they had caused significant damage
\citep[e.g.,][]{shaw2002puente}.

Given the accuracy and speed of our ROMs compared to the simulations, we
speculate that ROMs could complement empirical ground motion models used for
probabilistic seismic hazard analysis (PSHA) and earthquake early warning (EEW).
Recent earthquake simulations using southern California community velocity
and attenuation models, e.g., CVM-H \citep{suss2003p} and CVM-S \citep{magistrale1996geology}, can achieve good agreement with
real data, especially at longer periods \citep{lai2020shallow}. For example, simulations by \cite{hu20220} suggest
standard deviations of about 0.3 natural log units for 0.5 Hz Fourier amplitude
spectra data, compared to the \cite{bayless2019summary} GMM for
California which reports a total standard deviation of ~0.75 natural log units.
This suggests that our ROMs using southern California velocity
models should also achieve an adequate match with data
for real earthquakes. That said, a limitation in our approach is that our ROM will not accurately
extrapolate for sources located outside the box, $\mathcal{P}$ (see Fig. \ref{fig:map} and Fig \ref{fig:concept}).
Assuming a land area of 403,932 km$^2$ representing the state of California,
generating simulation data for 2-km spaced sources for a volume of
the entire state would require about 800,000 point source locations. This means
the total upfront computational cost would be $800,000\times6\times192$ CPUh = 921.6 million
CPUh. We expect that the approach in this paper should apply for other
areas that have 3D velocity models, such as the San Francisco Bay Area
\citep[e.g.,][]{hole2000three} or the Cascadia subduction zone \citep{stephenson2017p}.
Further validation by comparing ROMs against ground
motions recorded from real earthquakes would support
supplementing GMM predictions with physics-informed ROM
predictions in real-time earthquake alerts, though we save this validation for future work.
We note that applying these ROMs for source-based EEW \citep[e.g.,][]{bose2018finder,chung2019optimizing}
would also require determining a rapid focal mechanism solution and hypocentral depth.

For PSHA, our method allows us to obtain a surface
seismic wavefield with high spatial resolution because the method does not use reciprocity.
In comparison, the CyberShake framework \citep[e.g.,][]{graves2011cybershake}
uses the reciprocity property of the wave equation as they
evaluate suites of seismic timeseries for many different earthquake scenarios,
but a relatively limited number of sites. CyberShake then constructs probabilistic
seismic hazard maps by spatial interpolation of the site terms.
Thus, CyberShake simulations computationally scale with
the number of sites (generally on the order of hundreds), and not with the
total number of potential earthquake scenarios (usually on the order of
tens to hundreds of thousands of finite fault sources). To generate wavefields
with high spatial resolution with CyberShake, more sites would need to
be added and the computational cost would greatly increase.
However, with the ROM approach based on forward simulations,
adding more sites adds a negligible cost to the FOM simulations provided that the domain size is unchanged;
evaluating the ROM will become slightly more expensive with
each additional site due to the larger matrices that must be multiplied together
in Eq. \ref{eq:recon_elem}. However, by using this advantage, we speculate that ROMs could improve
the spatial resolution of probabilistic seismic hazard maps without drastically
increasing the computational requirements.

\section{Conclusion}
In conclusion, our study demonstrates the potential of
using reduced-order models to rapidly
and accurately generate physics-based seismic wavefields.
The method is based on interpolated proper orthogonal decomposition
using radial basis functions, and
generates three-component, velocity seismograms based on 3D wave
propagation simulations using a 3D seismic velocity model for southern California.
We find that the ROM approach offers a significant improvement in
computational efficiency, likely making it fast enough for real-time applications in
earthquake early warning and seismic hazard assessment.
Our method is also quite modular, in that it can be applied
for single general moment tensors as well as complex finite fault
rupture models composed of multiple subfaults with heterogeneous slip distributions and varying
source time functions. The speedup for elementary moment tensors is on the order
of $10^5$, and on the order of $10^2$ for a finite fault rupture
model consisting of 100 subfaults.
For an $M_W$ 5.0 earthquake, the mean absolute
error in the velocity seismograms is about 0.007 cm/s for the
horizontal components, and 0.004 cm/s for the vertical component.
These results confirm the
feasibility of using ROMs to approximate complex seismic wave dynamics
with high fidelity. Our results also provide a new perspective
on the interpolation of elastodynamic Green's functions, suggesting
that accurate interpolations can be obtained when the source-to-source
distance is up to $\approx$ 2 times the shortest wavelength.
This study shows progress towards providing rapid and accurate
Green's functions that might be used in future real-time applications and source inversions.

\section*{Acknowledgements}
We thank Wenyuan Fan for suggesting the usage of the moment tensor decomposition from \cite{kikuchi1991inversion}.
JMR acknowledges support from the National Science Foundation Graduate Research Fellowship Program (grant No. DGE-2038238).
DAM and AAG were supported by the National Science Foundation through grant Nos. EAR-2121568 and OAC-2311208.
Any opinions, findings, and conclusions or recommendations expressed in this material are those of the author(s) and do not necessarily reflect the views of the National Science Foundation.
We gratefully acknowledge the Gauss Center for Supercomputing e.V. (www.gauss-centre.eu)
for providing compute time on SuperMUC-NG, hosted at the Leibniz Supercomputing Center
(www.lrz.de), via project pn49ha.

\bibliographystyle{gji}
\bibliography{bib}

\appendix
\section{Elementary Moment Tensors}
\label{sec:append_mts}
The elementary moment tensors are fundamental components used to describe the seismic source mechanism.
A seismic source can be expressed as a combination of the six basis tensors $\widehat{\bm M}_i$.
The six elementary moment tensors are as follows \citep{kikuchi1991inversion}:

\begin{align*}
\widehat{\bm M}_1 &= \begin{bmatrix}
0 & 1 & 0 \\
1 & 0 & 0 \\
0 & 0 & 0
\end{bmatrix},
\qquad
\widehat{\bm M}_2 = \begin{bmatrix}
1 & 0 & 0 \\
0 & -1 & 0 \\
0 & 0 & 0
\end{bmatrix},
\qquad
\widehat{\bm M}_3 = \begin{bmatrix}
0 & 0 & 0 \\
0 & 0 & 1 \\
0 & 1 & 0
\end{bmatrix},
\\
\widehat{\bm M}_4 &= \begin{bmatrix}
0 & 0 & 1 \\
0 & 0 & 0 \\
1 & 0 & 0
\end{bmatrix},
\qquad
\widehat{\bm M}_5 = \begin{bmatrix}
-1 & 0 & 0 \\
0 & 0 & 0 \\
0 & 0 & 1
\end{bmatrix},
\qquad
\widehat{\bm M}_6 = \begin{bmatrix}
1 & 0 & 0 \\
0 & 1 & 0 \\
0 & 0 & 1
\end{bmatrix}.
\end{align*}

\end{document}